\documentclass[twocolumn]{aastex631}

\usepackage{booktabs}
\usepackage{mathtools}
\usepackage{tablefootnote}
\shorttitle{Call \& Response}
\shortauthors{Sellers, Milligan, \& McAteer}
\graphicspath{{./}}

\begin{document}

\received{February 21, 2022}
\revised{July 22, 2022}
\accepted{August 5, 2022}

\title{Call and Response: A Time-Resolved Study of Chromospheric Evaporation in a Large Solar Flare}

\author[0000-0001-5342-0701]{Sean G. Sellers}
\affiliation{Department of Astronomy, New Mexico State University MSC 4500 NM 88003-8001, USA}

\author[0000-0001-5031-1892]{Ryan O. Milligan}
\affiliation{Astrophysics Research Centre, School of Mathematics \& Physics, Queen's University Belfast University Road, Belfast, BT7 1NN, UK}

\author[0000-0003-1493-101X]{R.T. James McAteer}
\affiliation{Department of Astronomy, New Mexico State University MSC 4500 NM 88003-8001, USA}
\email{sellers@nmsu.edu, r.milligan@qub.ac.uk, mcateer@nmsu.edu}

\begin{abstract}
We studied an X1.6 solar flare produced by AR 12602 on 2014 October 22. The entirety of this event was covered by \textit{RHESSI}, \textit{IRIS}, and \textit{Hinode/EIS}, allowing analysis of the chromospheric response to a nonthermal electron driver. We derived the energy contained in nonthermal electrons via \textit{RHESSI} spectral fitting, and linked the time-dependent parameters of this call to the response in Doppler velocity, density, and nonthermal width across a broad temperature range. The total energy injected was $4.8\times10^{30}$~erg, and lasted $352$ seconds. This energy drove explosive chromospheric evaporation, with a delineation in both Doppler and nonthermal velocities at the flow reversal temperature, between 1.35--1.82~MK. The time of peak electron injection (14:06~UT) corresponded to the time of highest velocities. At this time, we found 200~km~s$^{-1}$ blueshifts in the core of \ion{Fe}{24}, which is typically assumed to be at rest. Shortly before this time, the nonthermal electron population had the shallowest spectral index ($\approx$ 6), corresponding to the peak nonthermal velocity in \ion{Si}{4} and \ion{Fe}{21}. Nonthermal velocities in \ion{Fe}{14}, formed near the flow reversal temperature were low, and not correlated with density or Doppler velocity. Nonthermal velocities in ions with similar temperatures were observed to increase and correlate with Doppler velocities, implying unresolved flows surrounding the flow reversal point. This study provides a comprehensive, time-resolved set of chromospheric diagnostics for a large X-class flare, along with a time-resolved energy injection profile, ideal for further modeling studies.
\end{abstract}

\section{Introduction} \label{sec:Intro}

Solar flares are considered a consequence of magnetic reconnection in the corona, resulting in the release of $\leq10^{32}$~erg over the course of the event. Approximately 20\% of the released energy is partitioned into the acceleration of particles in the corona \citep{Emslie2012}. A population of electrons is accelerated near the reconnection site to relativistic speeds, and stream down coronal loops away from the acceleration region \citep{Emslie2004,Emslie2012,Aschwanden2014}. The steep density increase in the transition region down to the chromosphere is generally responsible for the sudden deceleration of these accelerated particles. Energy is dissipated in the chromosphere primarily via Coulomb collisions, with a smaller amount of energy being dissipated via the bremsstrahlung process, which produces hard X-ray (HXR) emission via interaction with the ``thick-target'' chromosphere  \citep{Brown1971,Lin1976}. This energy injection in the chromosphere is likely the driving mechanism behind chromospheric evaporation, the process by which flares produce high-temperature, high-density plasma in the corona.
\par
The duration and evolution of HXR radiation varies dramatically from flare to flare. 
\cite{Warmuth2016} studied flares of several \textit{GOES} classifications, and found that the time ranges of significant nonthermal flux ranged from 0.4~minutes to over 50~minutes for longer events, but the analysis lacked information regarding the temporal evolution of the driving electron beam. Several studies attempted to resolve the time-dependence of the electron beam \citep{Kulinova2011,Kennedy2015,Fletcher2013}, and found the observed durations of the electron injection events to be on the order of several minutes. 
\cite{Holman2003} analyzed data from the Reuven Ramaty High Energy Solar Spectroscopic Imager (\textit{RHESSI}; \citealt{Lin2002}) of the X3.6 flare on 2002 July 23 and found nonthermal HXR emission lasting $\approx 10$ minutes.
\par

\begin{figure*}[ht]
    \centering
    \includegraphics[width = 18cm]{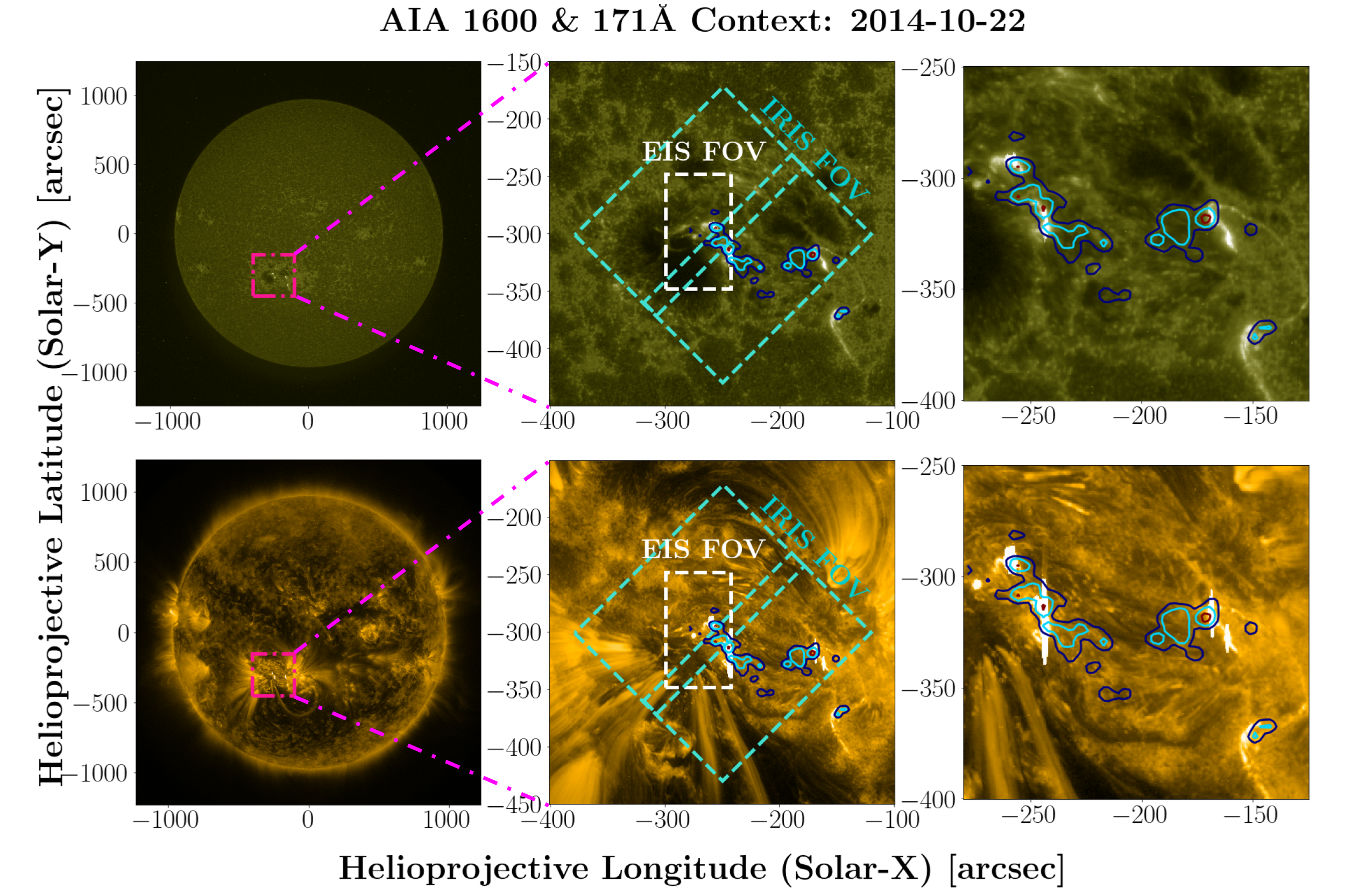}
    \caption{
    \textbf{Left:} Full-disk solar filtergrams on 2014 October 22 from \textit{AIA} 1600\AA\ (top) and 171\AA\ (bottom) bands.
    \textbf{Middle:} Image of NOAA 12192 during the X1.6 flare with \textit{RHESSI} 40--100 keV 20\%, 40\%, and 60\% contours (multiple colors), \textit{IRIS} slit-jaw imager and slit field of view (FOV) (cyan), and \textit{EIS} raster (white).
    \textbf{Right:} Detail of NOAA 12192 to highlight correlation between \textit{AIA} intensity enhancements and \textit{RHESSI} HXR footpoints.} 
    \label{fig:context}
\end{figure*}

\begin{figure*}[htb]
    \centering
    \includegraphics[width = 14cm]{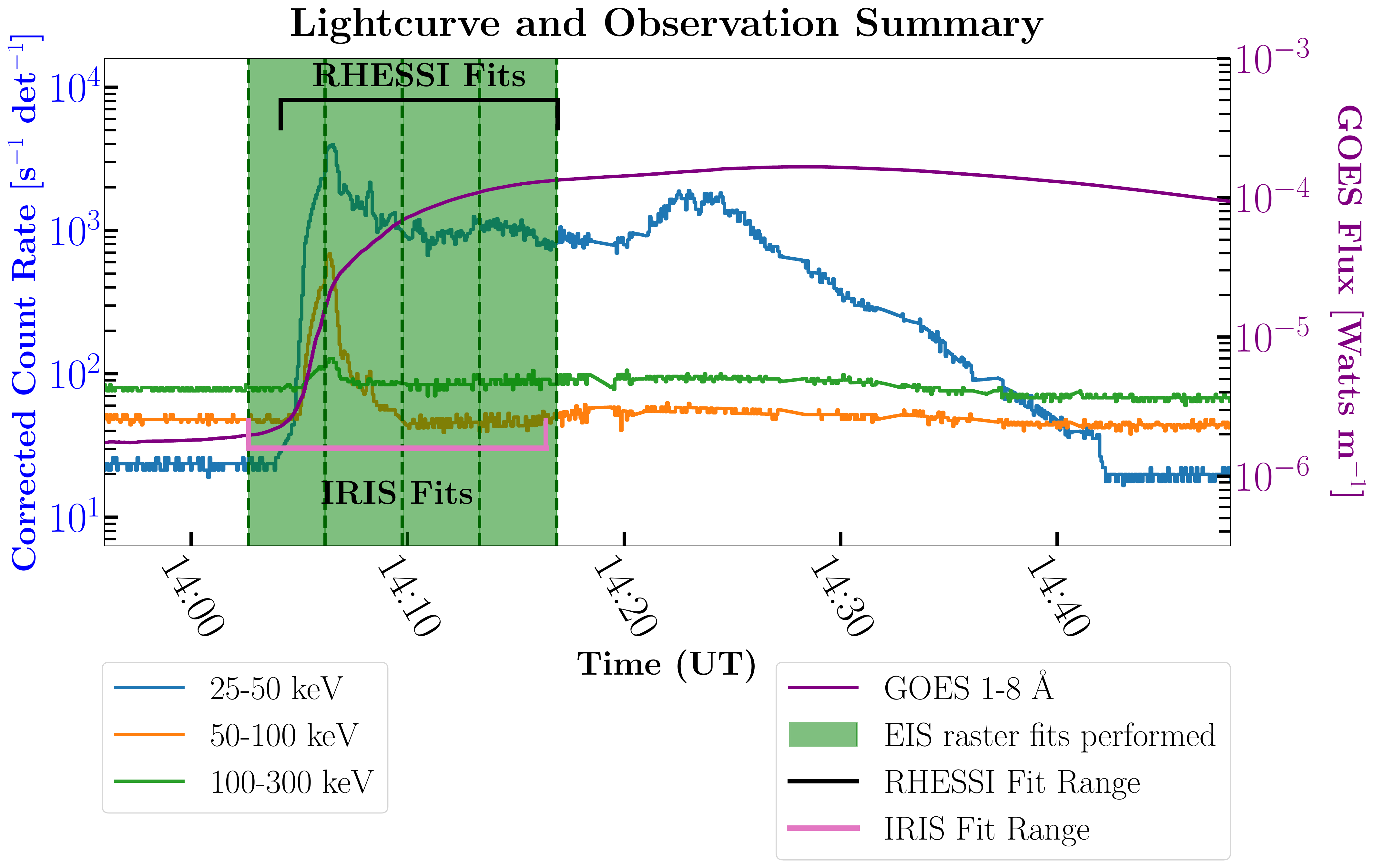}
    \caption{Lightcurves for \textit{GOES} (purple line) and \textit{RHESSI} (blue, orange, and green lines), with \textit{EIS} raster times overlaid (translucent green blocks), as well as the interval in which \textit{RHESSI} fits are performed (solid black line), and \textit{IRIS} fits are performed (solid pink line).}
    \label{fig:lc_context}
\end{figure*}

When the energy input to the chromosphere exceeds that which can be shed as radiation or conductive losses, the chromospheric plasma must heat and expand upward, into the lower-density corona. This process fills overlying magnetic structures of lower density with high-temperature plasma, which strongly emits extreme ultra-violet (EUV) and soft X-ray (SXR) emission. This chromospheric evaporation \citep{WernerM.Neupert1968,Bornmann1999,Fletcher2011} can occur explosively, with high-temperature lines exhibiting blueshifts, while cooler emission lines exhibit redshifts \citep{Doschek1983,Brosius2004,Milligan2009}; or gently, with blueshifted emission lines across a wide temperature range \citep{Fisher1985,Brosius2004,Allred2005,Milligan2006,Brosius2015}. The mode of evaporation is dependant first on the mechanism of flare energy transport. In the case of energy transport by a nonthermal electron driver, the mode of evaporation is further dependant on the energy flux, low energy cutoff, and population distribution of accelerated electrons reaching the chromospheric footpoints.
\par
\cite{Canfield1987} and \cite{Fisher1985} first deduced this effect and placed a lower limit on the requisite energy flux density required to drive explosive evaporation of $E_{e^-} \geq 3\times10^{10}$~erg~cm$^{-2}$~s$^{-1}$. If the incoming electron flux is above this threshold, determined by balancing the heating rate and the hydrodynamic expansion timescale, the over-pressure of the hot rising material causes the denser layers below to recoil, resulting in the cool, redshifted emission characteristic to explosive evaporation.
\par
Thermal conduction-driven chromospheric evaporation, in contrast, does not appear to be subject to above restrictions on flux deposition. \cite{Longcope2014} found that even the smallest energy fluxes studied produced explosive chromospheric evaporation. This result was also noted in earlier models from \cite{Fisher1989}.
\par
In addition to the Doppler velocity signatures of chromosperic evaporation, excess nonthermal width in optically thin spectral lines has been observed in flare conditions. One possible explanation is the superposition of unresolved flows. In this case, the nonthermal width is a measure of the velocity distribution of the plasma \citep{Doschek2008}. \cite{Newton1995} attempted to generalize both excess line widths and blue wing enhancements by the computation of a Velocity Differential Emission Measure (VDEM), which treats the observed line profile as a continuum of Gaussian components driven by variations in the line-of-sight velocity. This treatment is supported by reported correlations between Doppler velocities and nonthermal velocities within solar active regions \citep{Hara2008,Doschek2008,Bryans2010,Peter2010}. Another possible explanation for excess line widths is the influence of pressure or opacity broadening in regions of enhanced electron density. \cite{Milligan2011} showed a correlation between electron density and nonthermal velocity broadening, although neither pressure broadening nor opacity effects were able to account for any significant portion of the excess width.
\par
The flare-driven mass flow rate into the solar corona remains one of the more difficult solar flare metrics to disentangle from observations, requiring both accurate velocity information, and a measure of plasma mass. As a proxy, the electron density of the active region can be used \citep{Milligan2005,Doschek2008}. Density enhancements have been observed to be cospatial with the locations of flare footpoints \citep{Graham2011}. Densities, when combined with the emission measure \citep{DelZanna2011}, may also provide information about the dynamics of the evaporating region. The previously mentioned VDEM \citep{Newton1995} is derived in part from the electron density, and provides direct insight into plasma transport during a solar flare. 
\par
The flare chosen for the subject of this study, an X-class flare on 22 October, 2014, is a well studied event. \cite{Bamba2017} studied the precursor conditions to this event in order to determine triggering conditions in the chromosphere and photospheric magnetic field. \cite{Veronig2015} attempted to quantify the magnetic reconnection flux and rate. \cite{Li2015} utilized data from the \textit{Interface Region Imaging Spectrometer} (\textit{IRIS}; \citealt{IRISPaper}) and \textit{RHESSI} instruments to study Doppler velocities in \ion{Fe}{21} and \ion{C}{1}, and HXR intensities. \cite{Thalmann2015} focused on the rate of magnetic reconnection. \cite{Lee2017} measured electron flux at each HXR peak using \textit{RHESSI}, and linked the electron energy budget with observed low chromospheric and photospheric energetic response. These studies showed that energy was injected via high-energy electrons, which was sufficient to produce white-light emission.
\par
In this study, detailed, time-resolved \textit{RHESSI} HXR spectral fit parameters are presented in order to quantify the nonthermal electron energy injection profile. The profile of electron energy injection is then connected to multispectral observations of the chromospheric evaporation response. Emission line intensities, electron densities, and Doppler and nonthermal velocities from several instrumental sources were combined in order to study the response of the flaring solar atmosphere across time, space, and temperature. Due to the abundance of data available for this flare, this data set is ideally-suited to constrain detailed hydrodynamic modeling of energy transport during this event.
\par
An overview of this event, the data, and analysis techniques is presented in Section~\ref{sec:Analysis}. The results of this treatment and comparison to similar studies are discussed in Section~\ref{sec:Results}, and are summarized in Section~\ref{sec:Conclusions}. 

\section{Data Analysis} \label{sec:Analysis}

The X1.6 flare selected for study occurred on 2014 October 22, beginning at 14:02:00~UT, and was one of the largest flares produced by flare-productive NOAA AR 12192. In Figure~\ref{fig:context} the active region is presented in the 1600\AA\ and 171\AA\ passbands of the \textit{Solar Dynamics Observatory/Atmospheric Imaging Assembly} (\textit{SDO/AIA}; \citealt{Lemen2012}), with the fields of view of the \textit{EUV Imaging Spectrometer} (\textit{EIS}; \citealt{EISPaper}) and \textit{IRIS} instruments overlaid, and with HXR contours from \textit{RHESSI} imaging overlaid to highlight the primary footpoints of the flare. Two HXR sources are well-defined and are cospatial with intensity enhancements in \textit{AIA} images. A third, compact HXR kernel appears to the southwest of the primary flare loop, corresponding to a possible tertiary footpoint, or merely an extension of the large western footpoint. Figure~\ref{fig:lc_context} shows the \textit{RHESSI} HXR lightcurves in three energy bands (25--50, 50--100, and 100--300~keV) as well as SXR emission from the \textit{GOES} 1-8\AA\ band. Figure~\ref{fig:lc_context} provides additional context, with the time intervals where \textit{EIS}, \textit{IRIS} and \textit{RHESSI} spectral fits were performed. 
\par
The \textit{GOES} flux for this event plateaus through much of the event, with a SXR peak found well after the peak of HXR emission (14:28~UT, versus 14:06~UT). Hereafter, when the peak of the flare is referred to, it is in reference to the peak of HXR emission.

\subsection{RHESSI Analysis} \label{sec:RHESSI_anal}
\begin{figure}[t]
    \centering
    \includegraphics[width = \columnwidth]{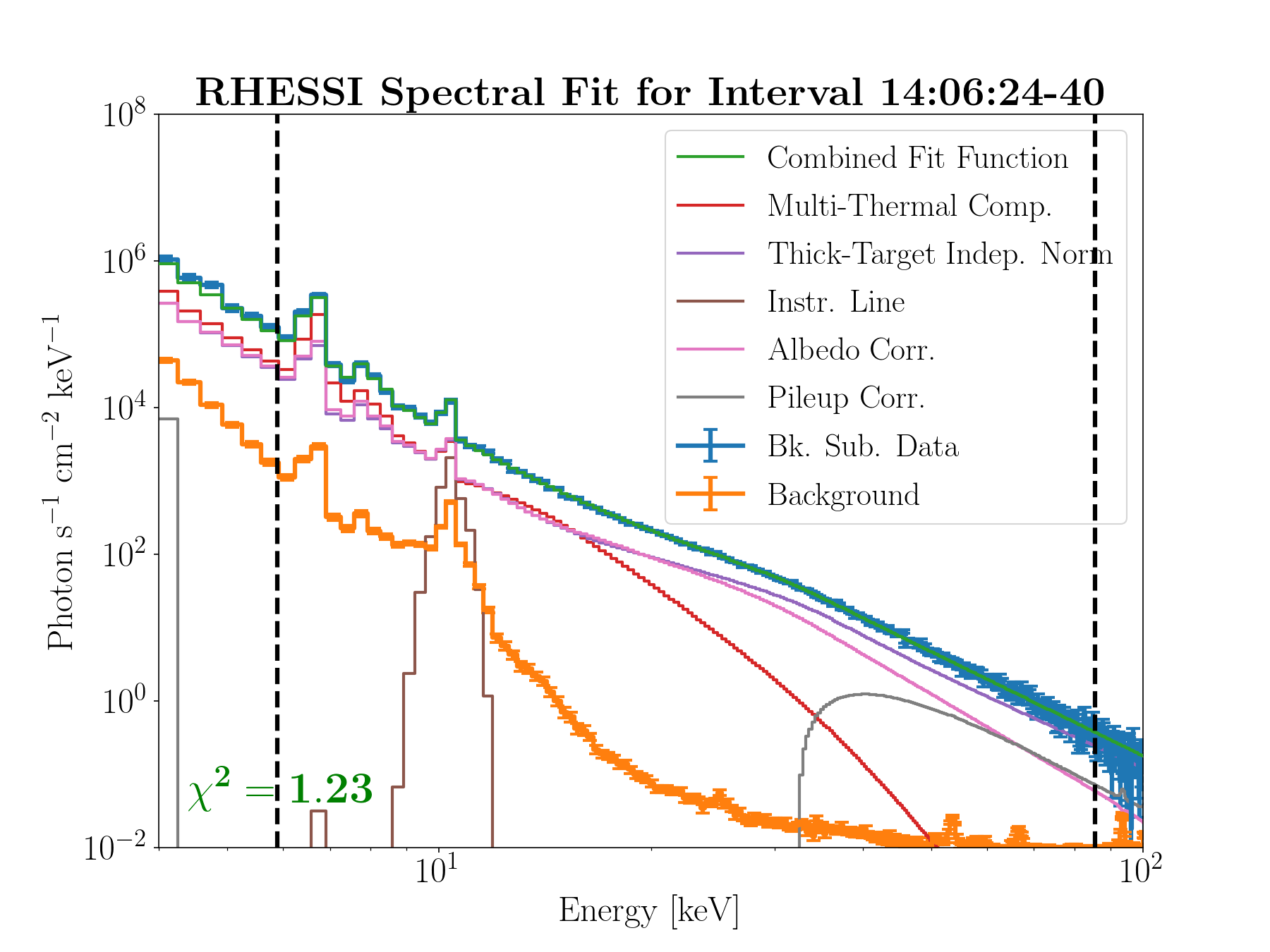}
    \includegraphics[width = \columnwidth]{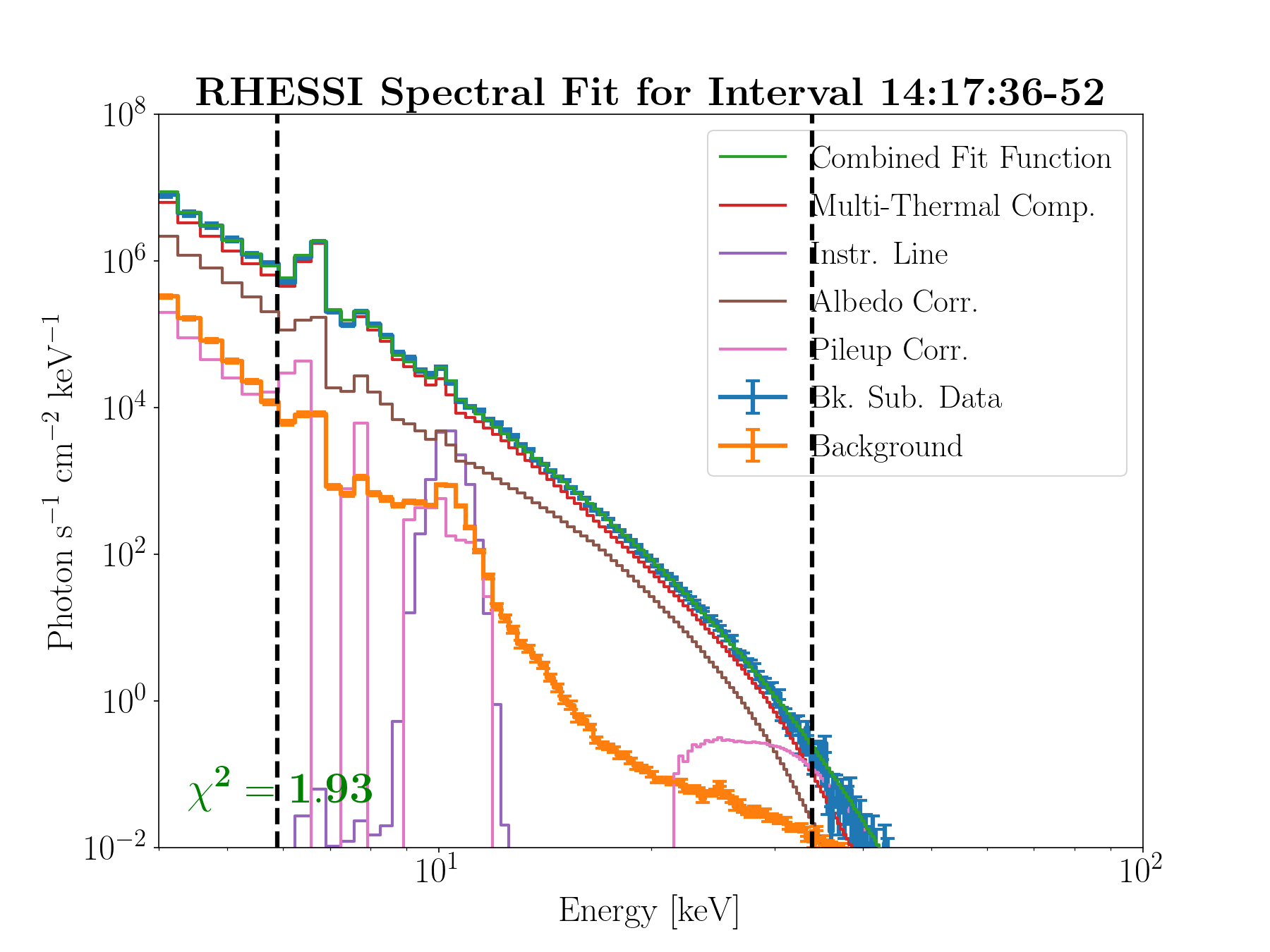}
    \caption{Example spectral fits from detector~$6$.
    \textbf{Top:} The 16-second interval with the highest integrated count level within the data set, from 14:06:24--14:06:40~UT. 
    The heavy black dashed line denotes the energy range fit. 
    \textbf{Bottom:} A sample 16-second interval after the cessation of nonthermal electron injection from 14:17:36--14:17:52~UT.}
    \label{fig:spex_fits}
\end{figure}

The full duration of this flare was well-covered by the \textit{RHESSI} instrument. \textit{RHESSI} entered its daylight phase just prior to the onset of the flare, and exited during the gradual phase, after the \textit{RHESSI} HXR peak and the \textit{GOES} SXR peak. As of August 2014, the \textit{RHESSI} spacecraft had undergone its fourth successful anneal, allowing five of the original nine detectors to regain high spectral resolution.
\par
\textit{RHESSI} spectra from 14:04:40 -- 14:16:56~UT were obtained with 16~second time bins for detectors 1, 3, 6, 8, and 9, which had consistently high count rates during the flare, signifying that they retained sufficient sensitivity to be usable. From the peak counts, detector 6 was determined to be the most sensitive, while detector 1 was the least, leading to different fit results and higher values of $\chi^2$ for detector 1. Background characterization and spectral fitting were performed for each individual detector using the \verb|OSPEX| package in \textit{SolarSoftWare} (SSW). For each detector, the background profile was determined by using the smoothed emission profile of the 100--300~keV energy band in the same detector. Save for one brief ($< 32$s) spike during the impulsive phase of the flare, emission in this energy range showed only a slow variation throughout the \textit{RHESSI} orbital cycle. This time-varying profile was used as a template for the background in lower energy bands. The count rate during \textit{RHESSI}'s night was used to determine the relative scaling between energy bands, and served as anchor points for application of the template.
\par
Spectra were fit using a methodology similar to that adopted by \cite{Milligan2014}. The thermal portion of the \textit{RHESSI} spectrum was best fit by a multithermal model, similar to studies by \cite{aschwanden2007}, \cite{Battaglia2015}, and \cite{Choithani2018}. The multithermal model selected was characterized by a power-law differential emission measure (DEM) between a fixed minimum plasma temperature (0.5~keV) and a variable maximum plasma temperature. The nonthermal portion of the \textit{RHESSI} spectrum was best fit by a thick-target electron beam model, with an electron distribution characterized by a single power-law. Additional instrumental effects were accounted for by modifying the detector response matrix (\verb|drm_mod|), accounting for instrumental pileup (\verb|pileup_mod|), albedo, and incorporating an additional Gaussian component to account for the 10~keV instrumental line \citep{Phillips2006}.
\par
Sample spectra are shown in Figure~\ref{fig:spex_fits} along with the combined fit functions used to characterize the HXR profile for a time interval with a significant nonthermal component (top panel) and a time interval without (bottom panel). Note that while spectra in Figure~\ref{fig:spex_fits} are shown in units of photons~s$^{-1}$~cm$^{-2}$~keV$^{-1}$, spectral fitting was carried out in count space. The use of the calculated photon spectrum exaggerates several notable features, such as the 10~keV instrumental line first characterized by \cite{Phillips2006}. Summaries of major parameters obtained via \textit{RHESSI} spectral fitting are discussed in Section~\ref{sec:hsi_results}.
\par
We also make use of the unique imaging capabilities of \textit{RHESSI} in order to identify the flare footpoints. The \verb|CLEAN| algorithm was applied to detectors 1, 3, 6, 8, and 9 during the impulsive and peak phases of the flare to identify sources of HXR emission throughout the flare duration. Contours of these images are overlaid on the center and right--hand columns of Figure~\ref{fig:context} to provide context for other observations and constrain the locations of HXR emission during the peak of the flare.

\subsection{EIS Analysis}\label{sec:EIS_anal}

\begin{figure*}
    \centering
    \includegraphics[width = 15cm]{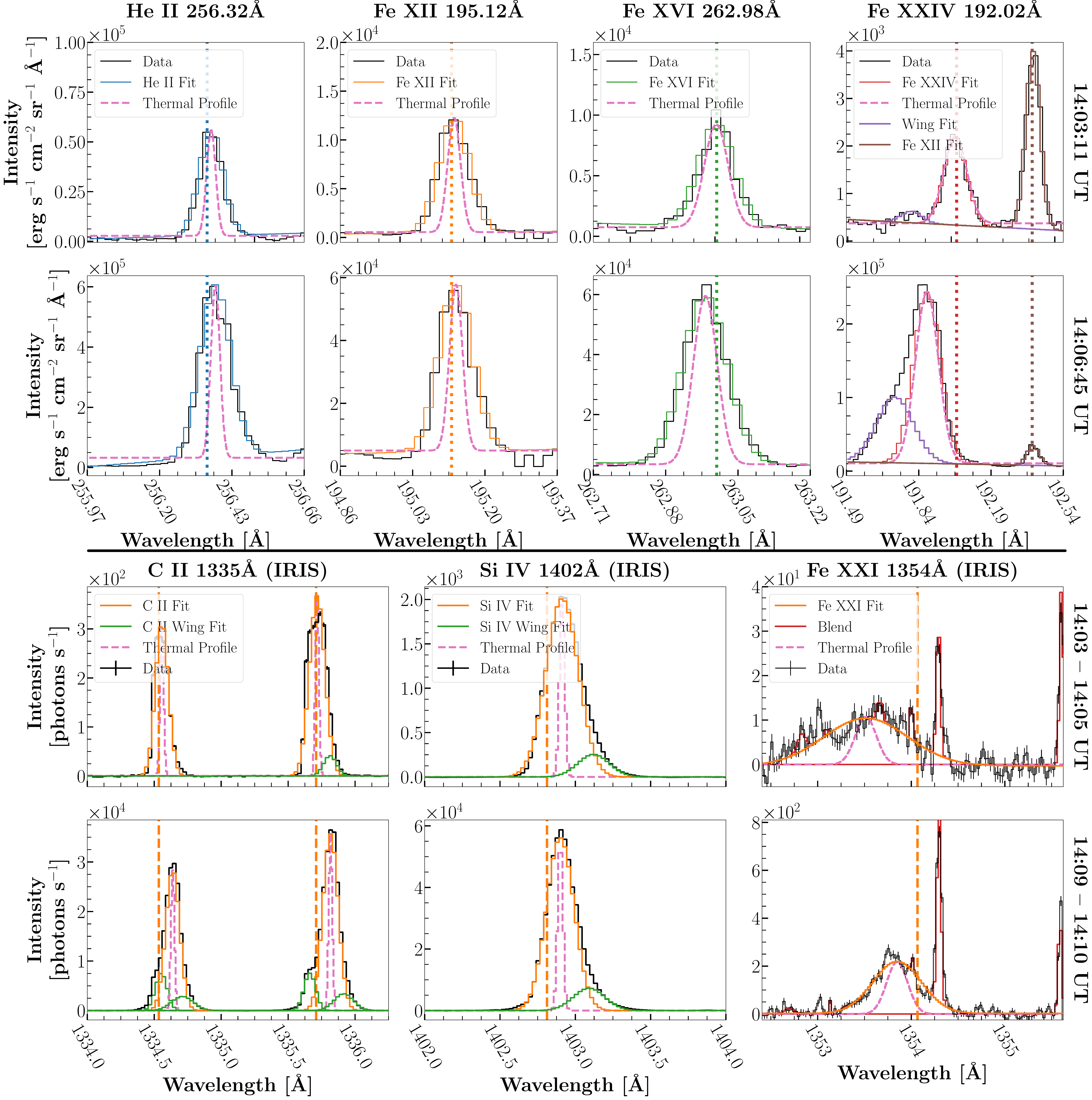}
    \caption{Example spectral fits from the \textit{EIS} and \textit{IRIS} instruments.
    The top two rows show fits from the \textit{EIS} instrument, while the bottom two rows show fits from the \textit{IRIS} instrument.
    For each instrument, the top row shows fits from a time early in the flare, while the bottom row shows fits from approximately the time of peak HXR emission.
    For the \textit{EIS} instrument, spectral windows containing \ion{He}{2} 256.35\AA , \ion{Fe}{12} 195.12\AA , \ion{Fe}{16} 262.98\AA , \ion{Fe}{24} 192.02\AA , and \ion{Fe}{12} 192.39\AA\ are shown left to right. 
    The faint blend with \ion{Fe}{12} 195.12\AA\ is not shown, as the major line dominates the window.
    For the \textit{IRIS} instrument, spectral windows containing the \ion{C}{2} 1334.54\AA\ and 1335.71\AA\ doublet, the \ion{Si}{4} 1402.81\AA\ line, and the \ion{Fe}{21} 1354.07\AA\ line are shown left to right.
    In each panel, the vertical line(s) denotes the calculated rest wavelength, while the pink dashed profile shows the profile of a line with the same peak and offset, but only thermal and instrumental width, without nonthermal broadening (Equation 1). 
    }
    \label{fig:eis_fit_ex}
\end{figure*}
\begin{table}
    \centering
    \caption{EIS and IRIS Line Summary}
    \begin{tabular}{p{0.15\linewidth}  p{0.25\linewidth}  p{0.4\linewidth}}
    \toprule
    \textbf{Ion} & \textbf{Formation Temperature [MK]$^{a}$} & \textbf{Central Wavelength (Angstrom)} \\
    \midrule
    \ion{Fe}{24} & 18.20 & $192.026 \pm 0.003$ \\
    \ion{Fe}{24} & 18.20 & $255.13 \pm 0.047$ \\
    \ion{Fe}{23} & 14.13 & $263.78 \pm 0.053$ \\
    \ion{Ca}{17} & 6.31 & $192.845 \pm 0.008$ \\
    \ion{Fe}{16} & 2.51 & $263.004 \pm 0.003$ \\
    \ion{Fe}{15} & 2.0 & $284.182 \pm 0.003$ \\
    \ion{Fe}{14} & 1.82 & $274.225 \pm 0.003$ \\
    \ion{Fe}{14} & 1.82 & $264.808 \pm 0.003$ \\
    \ion{Fe}{12} & 1.35 & $192.391 \pm 0.003$ \\
    \ion{Fe}{12} & 1.35 & $195.122 \pm 0.003$ \\
    \ion{Fe}{10} & 1.0 & $184.536 \pm 0.012$ \\
    \ion{He}{2} & 0.05 & $256.349 \pm 0.005$ \\
    \midrule
    \ion{Fe}{21} & 11.48 & $1354.067 \pm 0.04$ \\
    \ion{O}{1}$^{b}$ & N/A & $1355.599 \pm 0.04$ \\
    \ion{Si}{4} & 0.08 & $1402.812 \pm 0.057$ \\
    \ion{C}{2} & 0.01 & $1334.543 \pm 0.026$ \\
    \ion{C}{2} & 0.01 & $1335.705 \pm 0.024$ \\
    \bottomrule
    \end{tabular}
    \footnotesize{ $^a$Assuming ionization equilibrium.}
    \newline
    \footnotesize{ $^b$Used only for \ion{Fe}{21} reference wavelength}
    \label{tab:eis_ions}
\end{table}
\begin{figure}
    \centering
    \includegraphics[width = \columnwidth]{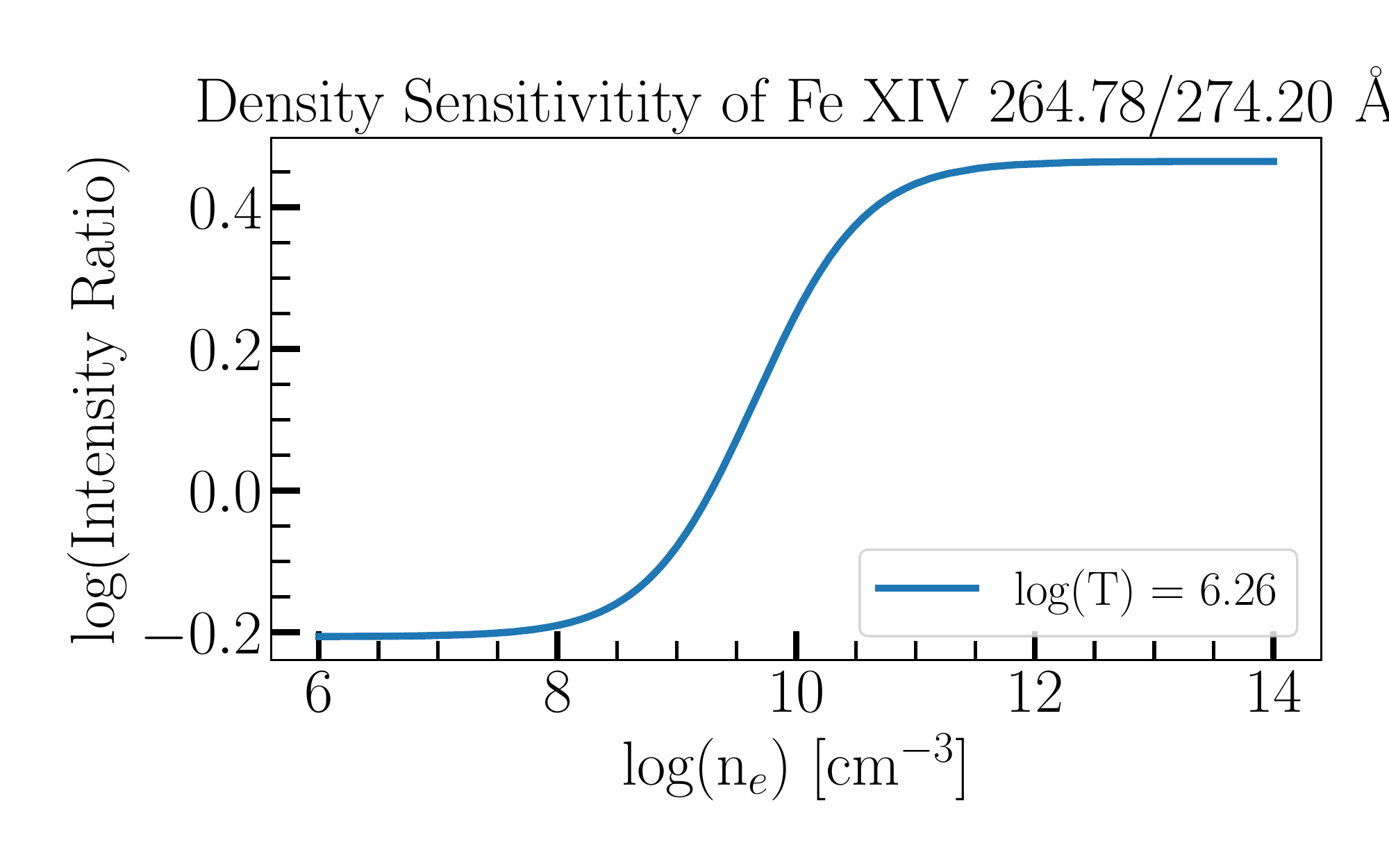}
    \caption{For \ion{Fe}{14} 264.81/274.23\AA\, the theoretical curve describing the relationship between the intensity ratio of each line, and the electron density \ion{Fe}{14}. 
    The theoretical curve is formed from the CHIANTI 10.0 atomic database \citep{Chianti1,Chianti2}.}
    \label{fig:dens_relation}
\end{figure}
Using the $2$\arcsec\ slit, the \textit{EIS} instrument performed rasters of NOAA AR~12192, capturing the pre-flare, impulsive, peak, and gradual phases of the solar flare centered around the eastern flare footpoint, as identified by \textit{RHESSI} HXR imaging. The rasters had a cadence of 214~seconds, covering a field of view (FOV) approximately 60\arcsec~$\times$~152\arcsec, as shown in Figure~\ref{fig:context}. The spatial resolution of the \textit{EIS} instrument is 3\arcsec\ in the horizontal, 1\arcsec\ in the vertical, with a spectral resolution of 22.3~m\AA. During this event, the footpoint was captured in several emission lines in the raster FOV; the observed emission lines are detailed in Table~\ref{tab:eis_ions}, which also includes information on emission lines from the \textit{IRIS} instrument. Gaussian fits were performed for a set of twelve emission lines from nine different ions. While most of the ions studied required only single-component fits, multiple component fits were performed in order to examine the effects of blended lines. The \ion{He}{2} 256.35\AA, \ion{Fe}{14} 272.20\AA, \ion{Fe}{15} 284.18\AA, and \ion{Ca}{17} 192.83\AA\ lines required two or more Gaussian profiles to account for known line blends \citep{Young2007}. Even in these spectral windows, the presence of strong blended lines was not consistent over each raster, or at each time. Additionally, the \ion{Fe}{23} 263.78\AA , \ion{Fe}{24} 255.13\AA , and \ion{Fe}{24} 192.02\AA\ lines required multiple components to account for both blends and a blue-wing enhancement \citep{Milligan2009}.
\par
The spectral fits were used to determine Doppler velocities, nonthermal velocities, electron densities, and intensities as functions of both temperature and time. Example fits from a selection of emission lines are shown in Figure~\ref{fig:eis_fit_ex}. The profiles chosen showcase a wide temperature range, from a location within the eastern footpoint early in the flare, and during the HXR peak.
\par
The rest wavelength for every emission line, save \ion{Fe}{23} and \ion{Fe}{24}, was determined from the mean central wavelength across the less-active raster regions. In the case of \ion{Fe}{23} and \ion{Fe}{24}, no plasma can be assumed to be at rest, and an alternate method was required. For \ion{Fe}{24} 192.02\AA\ the \ion{Fe}{12} 192.39\AA\ line was used to constrain the rest wavelength from the theoretical separation of the two lines from the CHIANTI database \citep{Chianti1,Chianti2}. For \ion{Fe}{23} and \ion{Fe}{24} 255.13\AA\ lines, the mean central wavelength from the 14:31:12~UT raster was used, as this raster consists entirely of emission produced after the nonthermal electron injection event. 
\par
For ions with strong blue wing enhancements (\ion{Fe}{23} and \ion{Fe}{24}), the Doppler velocities presented for the blue wing were calculated with the same reference wavelength used for the line core. \par
Nonthermal velocities were calculated using the method described by \cite{Mariska92}, and utilized in several other studies \citep{Doschek2007,Harra2009,Milligan2011}, where the most probable nonthermal velocity (v$_{nth}$) is calculated using the form:

\begin{equation}
    W^2 = 4 \ln{2} \Big(\frac{\lambda}{c}\Big)^2(v_{th}^2 + v_{nth}^2) + W_{inst}^2,
\end{equation}\label{eq:1}

where W is the measured full width at half maximum of the Gaussian profile, $W_{inst}$ is the instrumental width ($0.056$ m\AA\, \citealt{Doschek2007} and \citealt{Harra2009}). The thermal velocity, $v_{th}$ is given by

\begin{equation}
    v_{th} = \sqrt{\frac{2 k_B T}{M}},
\end{equation}

where $k_B$ is the Boltzmann constant, $M$ is the mass of the ion, and $T$ is the peak formation temperature of the line from \cite{Young2007}, and the CHIANTI database \citep{Chianti1,Chianti2}, assuming ionization equilibrium. 
\par
The \textit{EIS} dataset used in this work contains the density-sensitive line pair of \ion{Fe}{14} 264.81/274.23\AA. The theoretical relationship between the intensity ratio and electron density for this line pair is shown in Figure~\ref{fig:dens_relation}, from the CHIANTI v10.0 database \citep{Chianti1,Chianti2}. This line pair is sensitive to densities between $10^8 < n_e < 10^{12}$~cm$^{-3}$. It is important to note that the relationship between the \ion{Fe}{14} intensity ratio and electron density is formed under the assumption of ionization equilibrium, which may not be valid during large dynamic events such as solar flares. The Doppler and nonthermal velocity results from \textit{EIS} fitting are discussed in Sections~\ref{sec:cool_lines}~and~\ref{sec:hot_lines}, while the correlation between velocity parameters and electron density are discussed in Section~\ref{sec:eis_dens}.

\subsection{IRIS Analysis}\label{sec:IRIS_anal}

In this study, both the spectral and slit-jaw imaging data from the \textit{IRIS} instrument were used. For the entire duration of this event, the \textit{IRIS} instrument performed a repeated fast raster scan (131.1~s cadence per complete raster) of AR 12192, with a $45^{\circ}$ roll angle. Each spectral raster contained eight slit positions, with a spacing of 2\arcsec\ and 16.32~seconds between positions. The spatial resolution for each raster was 0.33\arcsec\ along the slit, with a slit width of 0.33\arcsec . No onboard spatial summing was carried out for these observations. The spectral resolution was 25.96~m\AA\ in the far-ultraviolet (FUV) spectral window.
\par
The \textit{IRIS} slit-jaw camera was used to determine the area of each flare footpoint. While this is not a direct measurement of the HXR source size, \textit{RHESSI} \verb|CLEAN| imaging tends to significantly overestimate the source size \citep{Dennis2009,Milligan2009}, and \textit{AIA} chromospheric images for this event were severely saturated during the period of interest.
\par
Ribbon areas were determined from \textit{IRIS} slit-jaw images using the 10\% and 50\% levels of each frame maximum. This time-dependant area measurement was interpolated from a 32-second cadence to a 16-second cadence, and rebinned to match the \textit{RHESSI} spectral time bins. The \textit{IRIS} slit-jaw camera experienced minimal saturation in two exposures during the peak of the flare; these were omitted from the final calculation of the footpoint area. The time-dependant areas were used to determine the injected electron energy flux in erg~s$^{-1}$~cm$^{-2}$.
\par
\textit{IRIS} spectra were available for several ion species during this flare, from which the \ion{C}{2} line doublets at 1334.54~and~1335.71\AA, the \ion{Si}{4} 1402.81\AA\ line, and the hot \ion{Fe}{21} line at 1354.07\AA\ were selected for study. Using the standard method described by \cite{IRIScal}, radiometric and instrumental calibrations were performed. The calibrated spectra were fit with multiple component Gaussian profiles, accounting for blends where applicable \citep{graham2015,young2015}, and allowing for additional blue- and red-wing components to account for asymmetry in the complex \ion{C}{2} and \ion{Si}{4} emission lines. 
\par
Despite the increased emissivity of the faint \ion{Fe}{21} line during the flare, it becomes more difficult to accurately fit during the peak of nonthermal electron injection. This is primarily due to the \textit{IRIS} instrument automatic exposure compensation, which scales exposures in order to avoid saturation in the more emissive ion species. During the peak of the flare, this has the unfortunate side effect of obscuring weak lines, such as \ion{Fe}{21}, within the noise of the continuum. In an attempt to maximize the signal from the \ion{Fe}{21} 1354.07\AA\ line, data from this spectral window were binned by a factor of four along the slit.
\par
Of the three species studied, only the \ion{Fe}{21} emission line is known to be optically thin. However, simulations have shown that Doppler shifts of the optically-thick \ion{C}{2} lines are well-correlated with the plasma velocity \citep{Rathore2015a,Rathore2015b,Rathore2015c}. The \ion{Si}{4} line is sometimes optically thin \citep{Kerr2019,cai2019,Peter2014}, with complex wavelength and structure-dependant behaviour \citep{Zhou2022}. Unfortunately, the diagnostic line at 1393\AA\ was not observed, and the opacity of the line could not be determined. Nevertheless, Doppler shifts were present within the line core, as were widths in excess of the thermal profile that could not be accounted for by known blends or observed asymmetry.  While the calculation of nonthermal velocity given by Equation~\ref{eq:1} is valid only for optically thin profiles, the same quantity calculated for an optically thick profile is a useful measure of line width. In the case of an optically thick line, variations in the width of the line are linked to changes in the optical depth of the line. As with the \textit{EIS} measurements, the quiescent regions in \ion{Si}{4} and \ion{C}{2} rasters were used to calculate reference rest wavelengths. For the broad \ion{Fe}{21} line, quiescent region emission of the nearby \ion{O}{1} line is used to infer the rest wavelength.

\section{Results}\label{sec:Results}

\subsection{RHESSI Results} \label{sec:hsi_results}

\begin{figure}[t]
    \centering
    \includegraphics[width = \columnwidth]{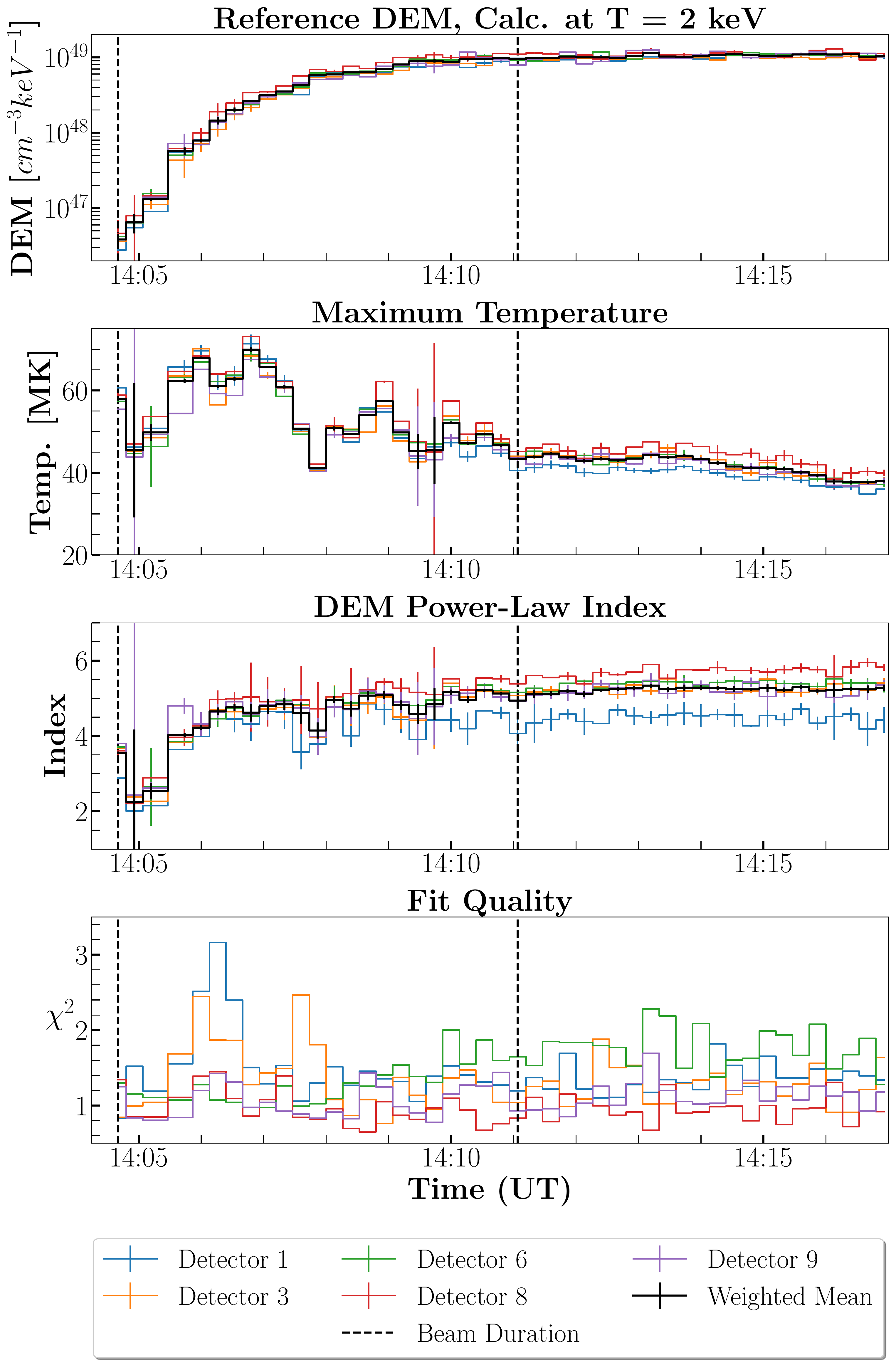}
    \caption{Thermal parameters from \textit{RHESSI} spectral fitting.
    \textbf{Panel 1:} The reference differential emission measure (DEM) calculated at T=2~keV $\approx 23.2$~MK) over all fitted intervals and detectors.
    \textbf{Panel 2:} The maximum temperature of the plasma in MK for all fit intervals and detectors.
    \textbf{Panel 3:} The power-law index used to calculate the DEM for all fit intervals and detectors.
    \textbf{Panel 4:} The reduced $\chi^2$ per fit interval and detector.
    The black dashed lines denote the range shown in Figure \ref{fig:hsi_nth}, where nonthermal emission was found.}
    \label{fig:hsi_therm}
\end{figure}

\textit{RHESSI} spectral fits were used to derive a set of thermal and nonthermal parameters for the X1.6-class flare on 2014 October 22. The thermal X-ray parameters, derived from the multithermal model are presented in Figure~\ref{fig:hsi_therm}. The top panel shows that the reference DEM (calculated at 2~keV, $\approx$ 23.2~MK) rose sharply soon after the onset of electron injection, and remained at approximately the same level ($\approx 10^{49}$~cm$^{-3}$~keV$^{-1}$), well after the cessation of the injection event. The upper limit on temperature, found in the second panel of the same figure, reached a peak of 70~MK early in the flare, and continued to decline for the rest of the studied interval. It is important, however, to note that this is the maximum temperature of the plasma, as characterized by a power-law DEM, and is not characteristic of the mean plasma temperature. The power-law index of the DEM increased slowly throughout the flare, as the bulk of the plasma cooled.
\par
\begin{figure}[t]
    \centering
    \includegraphics[width = 8cm]{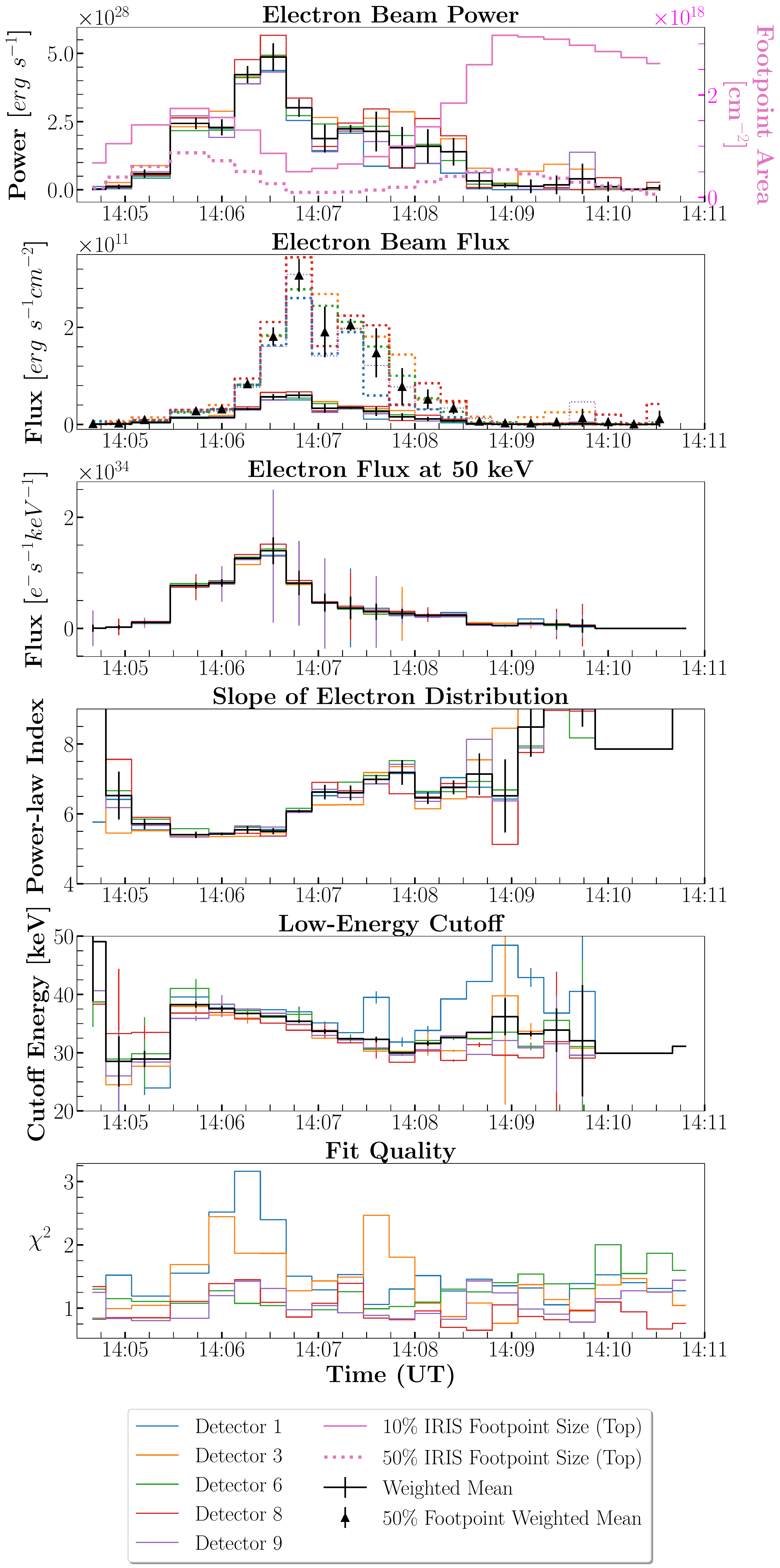}
    \caption{\textbf{Panel 1:} The power in nonthermal electrons from spectral fitting and two estimates for the \textit{IRIS} time-dependant footpoint area, as discussed in Section~\ref{sec:IRIS_anal}. The time-dependant footpoint areas are shown in pink, and correspond to the right-hand axis.
    \textbf{Panel 2:} The flux of nonthermal electrons, obtained by dividing the power shown in Panel 1 by the time-dependant footpoint area. The conservative estimate given by the 10\%\ footpoint contours is given as solid lines, while the estimate from a compact 50\%\ footpoint area is shown in the dotted lines.
    \textbf{Panel 3:} The electron flux at the reference energy of 50~keV in units of $10^{34}$~$e^-$~s$^{-1}$~keV$^{-1}$.
    \textbf{Panel 4:} The power law slope of the nonthermal electron distribution.
    \textbf{Panel 5:} The low-energy electron cutoff in keV.
    \textbf{Panel 6:} The reduced $\chi^2$ per fit interval and detector.}
    \label{fig:hsi_nth}
\end{figure}

The nonthermal electron parameters are presented in Figure~\ref{fig:hsi_nth}. The nonthermal electron population is best characterized by a single-power law distribution of electrons, that lasted for 352 seconds, and deposited $>4.8\times 10^{30}$~erg of energy. The nonthermal electron flux was first observed during the 14:04:40--14:04:56~UT interval, peaked during the interval 14:06:40--14:06:56~UT, 68~seconds after the first interval where the presence of nonthermal electrons was detected, and had ceased by 14:10:32~UT. 
\par
During the peak interval, the flux in nonthermal electrons was calculated to be between $5.99 \pm 0.66 \times10^{10}$~erg~s$^{-1}$~cm$^{-2}$, for a larger estimate of the footpoint area (corresponding to 10\% of the frame maximum for \textit{IRIS} slit-jaw imaging) and $3.07\pm0.34\times10^{11}$~erg~s$^{-1}$~cm$^{-2}$, for a smaller estimate of the footpoint area (the 50\% of the frame maximum).
\par
\cite{Lee2017} fit the \textit{RHESSI} spectrum of this event for two intervals during this flare, and calculated an energy flux of $7.7\times10^{10}$~erg~cm$^{-2}$~s$^{-1}$ during the time interval 14:05:32--14:06:32~UT. This is similar to the value obtained for the time interval 14:06:16--14:06:32~UT of $8.37 \pm 0.62 \times 10^{10}$~erg~cm$^{-2}$~s$^{-1}$ for the more conservative 50\% intensity threshold used to determine the area of the energy injection region. These results presented here are not compared with results from the second interval shown by \cite{Lee2017} ($6.1\times10^{10}$~erg~cm$^{-2}$~s$^{-1}$ at 14:11~UT). The differences between the these two studies are primarily due to differences in footpoint area determination and the determination of the low-energy electron cutoff. This study used the time-varying 10\% and 50\% contours of \textit{IRIS} imaging for footpoint area determination while \cite{Lee2017} take the 60\%\ contour of \textit{RHESSI} HXR imaging. This study additionally allows the low-energy electron cutoff to vary in time. This results in a cutoff between 5 and 8~keV higher than the 30~keV assumed by \cite{Lee2017}. The treatment presented here additionally fits for albedo effects and instrumental pileup. 
\par

In general, the low-energy cutoff presented in this work was higher than found in other, similar, studies, particularly \cite{Milligan2014}, who studied a flare of a similar size (X2.2). The study by \cite{Warmuth2016} contained several flares of similar magnitude, all of which had low-energy cutoffs less than found here. Most similar was the X1.3 flare of 2005 January 19, studied by \cite{Warmuth2009}, who found a low-energy cutoff between 30-40~keV during parts of that event. Due to the low-energy electron electron cutoff level, the particularly steep slope of nonthermal emission, and the choice of a multithermal plasma model, the derived electron power was, on the whole, weaker than studies of flares of a similar size. As with other studies \citep{xia2021}, a consequence of uncertainty in the low-energy cutoff is that the $4.8\times10^{30}$~erg total energy fit should be treated only as a lower limit \citep{Warmuth2009,aschwanden2019}.
\par

In Figure~\ref{fig:lc_context}, a secondary enhancement in the \textit{RHESSI} 25--50~keV band occurred around 14:24~UT. At the same time, flux in the \textit{GOES} 1--8\AA\ band is boosted. Taken together, this would imply the existence of a second nonthermal event at this time. Fits to the HXR spectrum were attempted from 14:15~UT through this secondary peak, until 14:30~UT, but the presence of a second nonthermal event could not be determined. Excess HXR emission was equally well fit by a thick-target bremsstrahlung component as by a pulse-pileup phenomenon component, with both cases yielding a similar $\chi^2$. \verb|CLEAN| images formed during this interval showed no significant sources of emission above the 30~keV low-energy electron cutoff derived during the nonthermal electron event.

\subsection{EIS Results} \label{sec:eis_results}
\subsubsection{Lines formed below 10~MK}\label{sec:cool_lines}

\begin{figure*}
    \centering
    \includegraphics[width = 12.4cm]{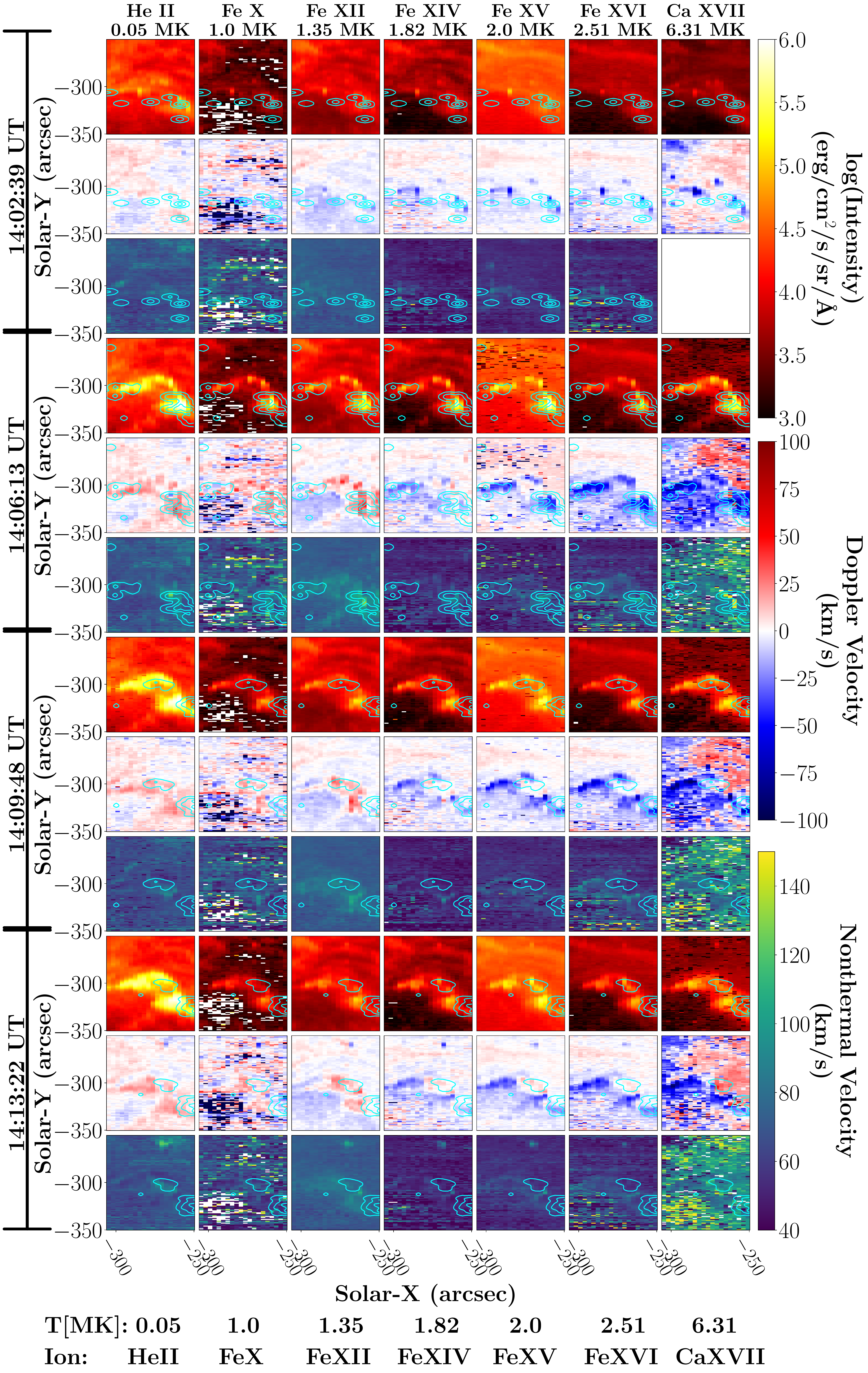}
    \caption{Line intensities, Doppler velocities, and nonthermal velocities for ions with $T < 10$ MK, for four rasters beginning at: 14:02:39~UT, 14:06:13~UT, 14:09:48~UT, and 14:13:22~UT, corresponding approximately to the pre-flare/early impulsive phase, peak impulsive phase, end of impulsive/beginning of the gradual phase, and after the cessation of nonthermal electron injection, respectively. Each column represents a different ion, with formation temperatures increasing left-to-right across the figure. The provided scales for each parameter are scaled to be consistent between observed ions and observation times. In the case of \ion{Ca}{17}, there was no significant nonthermal velocity at 14:02:39~UT, and this panel is left blank. \textit{RHESSI} 25--50~keV CLEAN imaging contours are overlaid in cyan on all images, corresponding to 5\%, 30\%, and 70\% of the image maximum.}
    \label{fig:eis_cool_ions}
\end{figure*}

Fit-derived parameters from ions with temperature T~$<$~10~MK are shown in Figure~\ref{fig:eis_cool_ions} for the four rasters spanning 14:02:39--14:16:56~UT. Line intensities, Doppler velocities, and nonthermal velocities are shown as rows in Figure~\ref{fig:eis_cool_ions} for each raster time interval, with ion formation temperature increasing left to right across each row. Columns in Figure~\ref{fig:eis_cool_ions} correspond to one emission line each (labelled at the top of each column). Each parameter was scaled to the same range across each time interval and temperature, to allow direct comparison between ion species, and the location of the flare footpoint (from \textit{RHESSI} 25--50~keV \verb|CLEAN| images) is overlaid in cyan. All HXR sources are part of the eastern flare footpoint; the western footpoint lay outside the \textit{EIS} FOV. Alignment between \textit{EIS} rasters and \textit{RHESSI} imaging was performed by first determining the offset between \textit{EIS} rasters and \textit{AIA} filtergrams using the \verb|eis_aia_offsets| procedure available in SSW, then aligning \textit{AIA} filtergrams with \textit{RHESSI} \verb|CLEAN| maps. The alignment between \textit{AIA} and \textit{EIS} is accurate to within $\approx 5$\arcsec \citep{EISSW}. The accuracy of alignment between \textit{AIA} and \textit{RHESSI} is accurate to $2.26$\arcsec, within the minimum spatial resolution element of \textit{RHESSI} imaging. The middle column of Figure~\ref{fig:context} shows the \textit{EIS} FOV in context of the flaring region for comparison to the structures shown in Figure~\ref{fig:eis_cool_ions}.
\par
In the rasters beginning at 14:06:13,~UT, 14:09:48~UT, and 14:13:22~UT, the velocity distribution found in \textit{EIS} spectral lines was typical of explosive chromospheric evaporation. Within the flare footpoint, warmer ions exhibited strong blueshifts, while cooler ions exhibited only redshifts. Given adequate temperature sampling, the Doppler velocities \textit{EIS} spectral lines can be used to derive a range for the temperature of flow reversal. The flow reversal temperature (FRT) is the temperature at which the division between evaporative upflows and condensation-driven downflows occurs during periods of explosive chromospheric evaporation. Analysis of Doppler velocities at or near this temperature provide insight into the processes that transport energy from the corona to the chromosphere \citep{Brannon2014,Fisher1985}. With six different ion species between T=1~MK and T=6.3~MK, the \textit{EIS} observations presented in this study are adequate to place constraints on this temperature.
\par
Figure~\ref{fig:eis_cool_ions} shows a clear delineation in Doppler velocity cospatial with HXR emission between 1.35--1.82~MK, first observed in the 14:06:13~UT raster. This raster spanned the time interval with the largest nonthermal electron flux density (Figure~\ref{fig:hsi_nth}). The distribution of nonthermal electrons during this interval was characterized by a steepening power-law index. In this, and the two following rasters, the \ion{Fe}{12} line, formed at 1.35~MK, exhibited mild downflows within the flare footpoint, on the order of $\approx10$--$40$~km~s$^{-1}$, while the \ion{Fe}{14} line, formed at 1.82~MK, was blueshifted between $\approx -20$ and $\approx$ $-60$~km~s$^{-1}$. The FRT fell within this 0.5~MK range during this raster, and remained in this range for the remainder of the flare. This range is consistent with limits determined in previous studies \citep{Kamio2005,Milligan2009}. Above this temperature, spectral lines were observed to have increasingly strong blueshifts, peaking at nearly -100~km~$^{-1}$ for the \ion{Ca}{17} line, while the cooler ions exhibited relatively consistent redshifted emission across the three cool species studied, including the weak \ion{Fe}{10} line.
\par
Minor evolution in the Doppler velocity distribution was found throughout the duration of the flare. The earliest raster studied, which began at 14:02:39~UT, showed markedly different behaviour compared to later observations. Nonthermal emission from \textit{RHESSI} observations were first observed at 14:04:40~UT, thus, this raster observed both the pre-flare and early-flare chromosphere. As early as 14:03:00--14:03:11~UT, ions warmer than the FRT were observed to have blueshifted velocity enhancements, 90~s before the \textit{RHESSI} instrument detected nonthermal emission. The \ion{Fe}{16} ion, in particular displayed a blueshift of -68.9$\pm$4.6~km~s$^{-1}$ in the region that subsequently became the flare footpoint.
This early velocity behaviour is more consistent with gentle chromospheric evaporation \citep{Schmieder87,Zarro88}, possibly driven by a nonthermal electron component with an energy below the \textit{RHESSI} sensitivity threshold.
\par
The compact kernels of blueshifted emission apparent in warm ($\geq$1.82~MK) ions during the 14:02:39~UT raster expanded to fill both lobes of the flare ribbon during the 14:06:13~UT raster. At this time, additional blueshifted material bridged the two HXR sources.
By the 14:13:22~UT raster, while significant upflows remained in these species, they were mostly contained within the eastern structure, while the larger, western structure had begun to return to rest as early as the 14:09:48~UT raster. The blueshifted material bridging the two HXR sources persisted through the 14:09:48~UT raster, but is largely absent by 14:13:22~UT.
\par
During the 14:02:39~UT raster, ions cooler than the FRT (\ion{He}{2}, \ion{Fe}{10}, and \ion{Fe}{12}) exhibited small Doppler velocity enhancements within the region that would later become the flare footpoint. The downflows in these species peak during the 14:06:13~UT raster (for \ion{He}{2}, downflows peaked at $v_{max }=$~41.7~$\pm$~5.5~km~s$^{-1}$ during this raster), gradually returning to rest over the remaining duration. The results presented here are broadly consistent with the results of \cite{Lee2017}, who presented selected ion species within a point in the western lobe.
\par
Nonthermal velocities (calculated from the line width) are shown in every third tow of Figure~\ref{fig:eis_cool_ions}. The highest nonthermal velocities derived from \textit{EIS} spectral fits were found at cooler temperatures, specifically, those below the FRT, and are largest for \ion{Fe}{12} and \ion{He}{2}. The ion observed by \textit{EIS} with the smallest nonthermal velocity was \ion{Fe}{14}, which is formed at a temperature just above the FRT. Ions warmer than \ion{Fe}{14} showed higher nonthermal velocities with increasing temperature. There is little evolution in nonthermal velocity after 14:06:13~UT. During the 14:02:39~UT raster, the nonthermal velocity, particularly in \ion{Fe}{14}, \ion{Fe}{15}, and \ion{Fe}{16} was mildly enhanced across the region that would later become the flare footpoint. Overall, the nonthermal velocities observed are markedly similar in magnitude to those observed by \cite{Milligan2011}, though the flare studied in that work was significantly smaller (C1.1).

\subsubsection{Lines formed above 10~MK}\label{sec:hot_lines}
\begin{figure*}
    \centering
    \includegraphics[width=13.5cm]{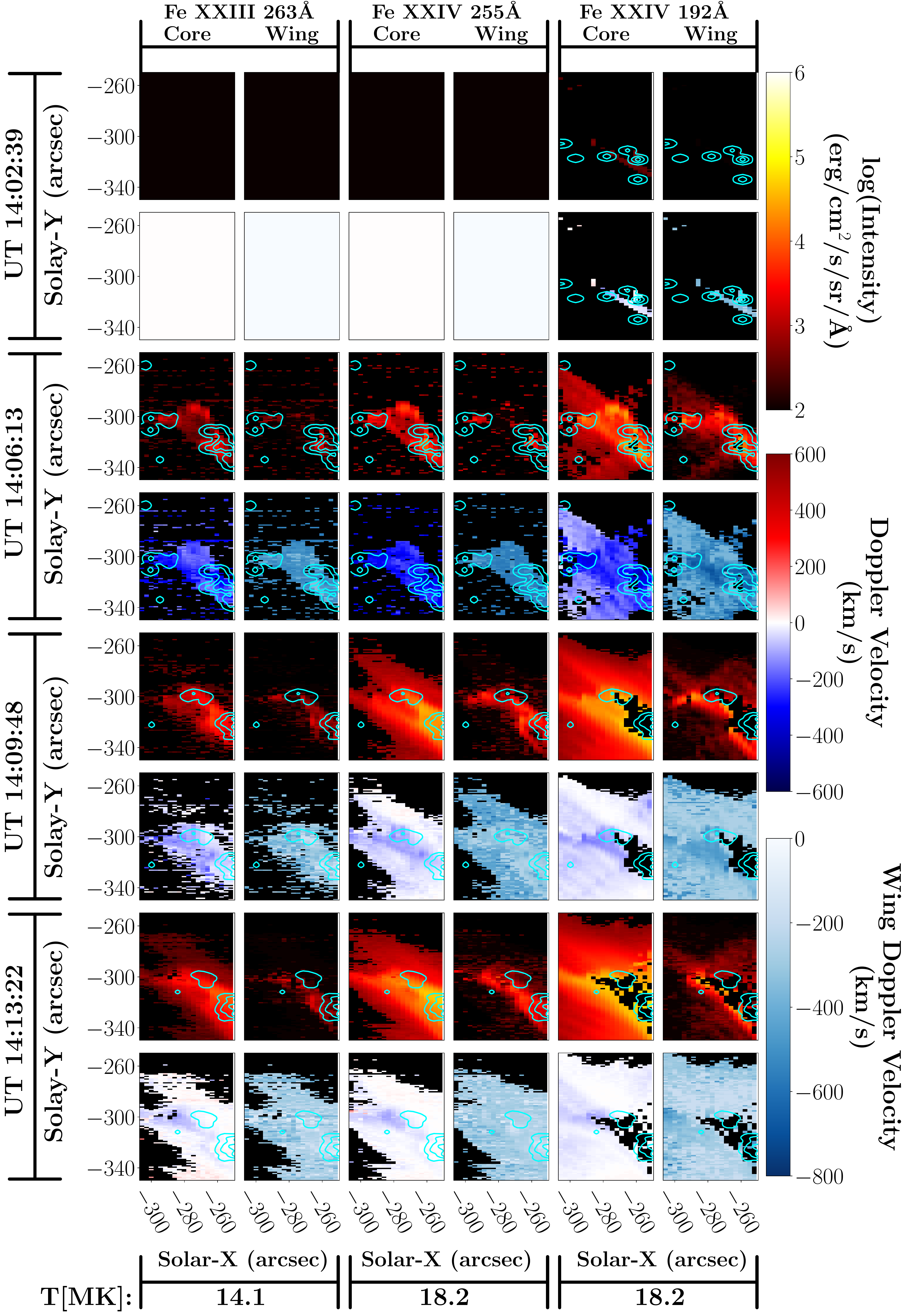}
    \caption{Line intensities, line core Doppler velocities, and velocity separation of blue-wing components for ions \ion{Fe}{23} 263.78\AA, \ion{Fe}{24} 255.13\AA, and \ion{Fe}{24} 192.02\AA, for four rasters beginning at 14:02:39~UT, 14:06:13~UT, 14:09:48~UT, and 14:13:22~UT. These times correspond approximately to the pre-flare/early impulsive phase, peak impulsive phase, end of impulsive/beginning of the gradual phase, and after the cessation of nonthermal electron injection. Each raster time occupies two rows (intensity and Doppler velocity for each time interval), while each ion occupies two columns (line core and blue wing). 
    The colorbar scale is consistent between ion species and observation times. 
    At the earliest times, only \ion{Fe}{24} 192.02\AA\ displays emission above the background. 
    \textit{RHESSI} 25--50~keV CLEAN imaging contours are overlaid in cyan on all images, corresponding to 5\%, 30\%, and 70\% of the image maximum.}
    \label{fig:eis_hot_ions}
\end{figure*}

Figure~\ref{fig:eis_hot_ions} shows Doppler velocity behaviour for the hottest \textit{EIS} ions: \ion{Fe}{23} and \ion{Fe}{24}, with temperatures of 14.13~and~18.20~MK respectively. The \textit{EIS} instrument observed one line of \ion{Fe}{23}, and two of \ion{Fe}{24}. Of the three, the reference wavelength constraints for \ion{Fe}{24} 192.02\AA\ were most reliable, and this line serves as the focus of this discussion. In every raster where these lines were present, the core was accompanied by strong enhancements to the blue wing. Figure~\ref{fig:eis_hot_ions} presents the fits to the core and the blue wing enhancement, with time increasing top to bottom for the same four raster time intervals presented in Figure~\ref{fig:eis_cool_ions}. The six columns correspond to: \ion{Fe}{23} core and blue wing, the \ion{Fe}{24} 255.13\AA\ core and blue wing, and the \ion{Fe}{24} 192.02\AA\ core and blue wing, while the rows alternate between intensity and Doppler velocity for these four components. For the blue wing, Doppler velocity was measured relative to the same reference wavelength as the line core. Where the detector saturated observing \ion{Fe}{24} 192.02\AA, or where there was insufficient signal to fit the emission line, as was often the case outside the flare ribbon, the fits were replaced with a null value.
\par
During the 14:02:39~UT raster, no emission was detected from the \ion{Fe}{23} line or the \ion{Fe}{24} 255.13\AA\ line. The \ion{Fe}{24} 192.02\AA\ line, while faint, was present in locations that later became a part of the footpoint during this time interval. 
\par
\begin{figure}
    \centering
    \includegraphics[width = \columnwidth]{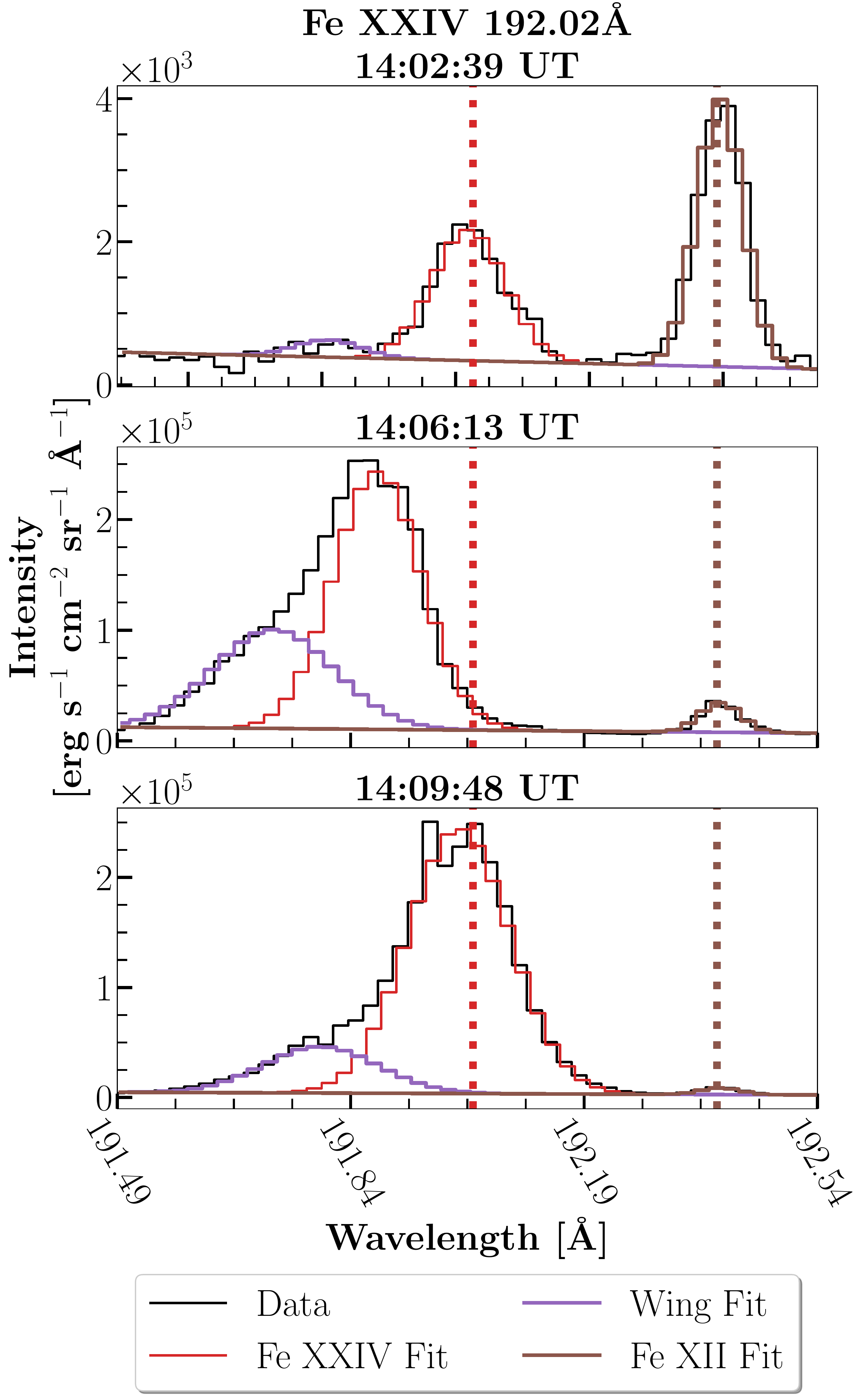}
    \caption{The \ion{Fe}{24} 192.02\AA\ complex for three times, 14:02:39~UT (\textbf{Top}, pre-nonthermal HXR), 14:06:13~UT (\textbf{Middle}, peak nonthermal HXR), and 14:09:48~UT (\textbf{Bottom}, late nonthermal HXR). 
    Fit profiles for the core of the line are shown along with the fit to the blue wing enhancement and the fit to the \ion{Fe}{12} 192.39\AA\ line used for reference wavelength determination. 
    The rest position of \ion{Fe}{24} is overlaid as the dotted red line, and the dotted brown line denotes the rest position of the \ion{Fe}{12} line. 
    Note the difference in intensity scale between the top and middle panels.}
    \label{fig:fexxiv_complex_shift}
\end{figure}
An example of this early, low-intensity emission is shown in the top panel of Figure~\ref{fig:fexxiv_complex_shift}. Where it is present at 14:02:39~UT, the magnitude of Doppler velocity for \ion{Fe}{24} is small for the core, and the separation of the wing is approximately constant.
\par
Significant \ion{Fe}{23} and \ion{Fe}{24} 255.13\AA\ emission first appeared during the 14:06:13~UT raster, and grew in intensity with each successive raster. All three lines exhibited core blueshifts within the footpoint at this time, with further blue-wing enhancement.
The Doppler velocity of the blue wing peaked during the 14:06:13~UT raster, and decreased thereafter.
\par
Generally, these hot ions are expected to display a stationary core, with an enhanced blue wing \citep{Milligan2009}. During the raster covering the flare peak (14:06:13~UT), however, the entire line complex for both the \ion{Fe}{23} line, and the \ion{Fe}{24} line pair was significantly blueshifted. Within non-saturated footpoint pixels, the \textit{core} of the \ion{Fe}{24} 192.02\AA line was found to have blueshifts as high as -240~km~s$^{-1}$, while maintaining a blue wing enhancement. For the same profile, the blue wing velocity was as high as -480~km~s$^{-1}$, relative to the same reference wavelength. By the 19:09:48~UT raster, while core blueshifts were still found within the flare ribbon, the magnitude and extent were far less than found one raster prior, and by 14:13:22~UT the core of these lines had mostly returned to rest. Significant Doppler velocities observed in the ``rest'' component of this line complex is not expected. An example of this atypical behaviour is shown in Figure~\ref{fig:fexxiv_complex_shift}, which shows the \ion{Fe}{24} 192.02\AA complex across three rasters from the same location.

\subsubsection{Correlations between Doppler and Nonthermal Velocity, and Electron Density}\label{sec:eis_dens}

\begin{figure}
    \centering
    \includegraphics[width = \columnwidth]{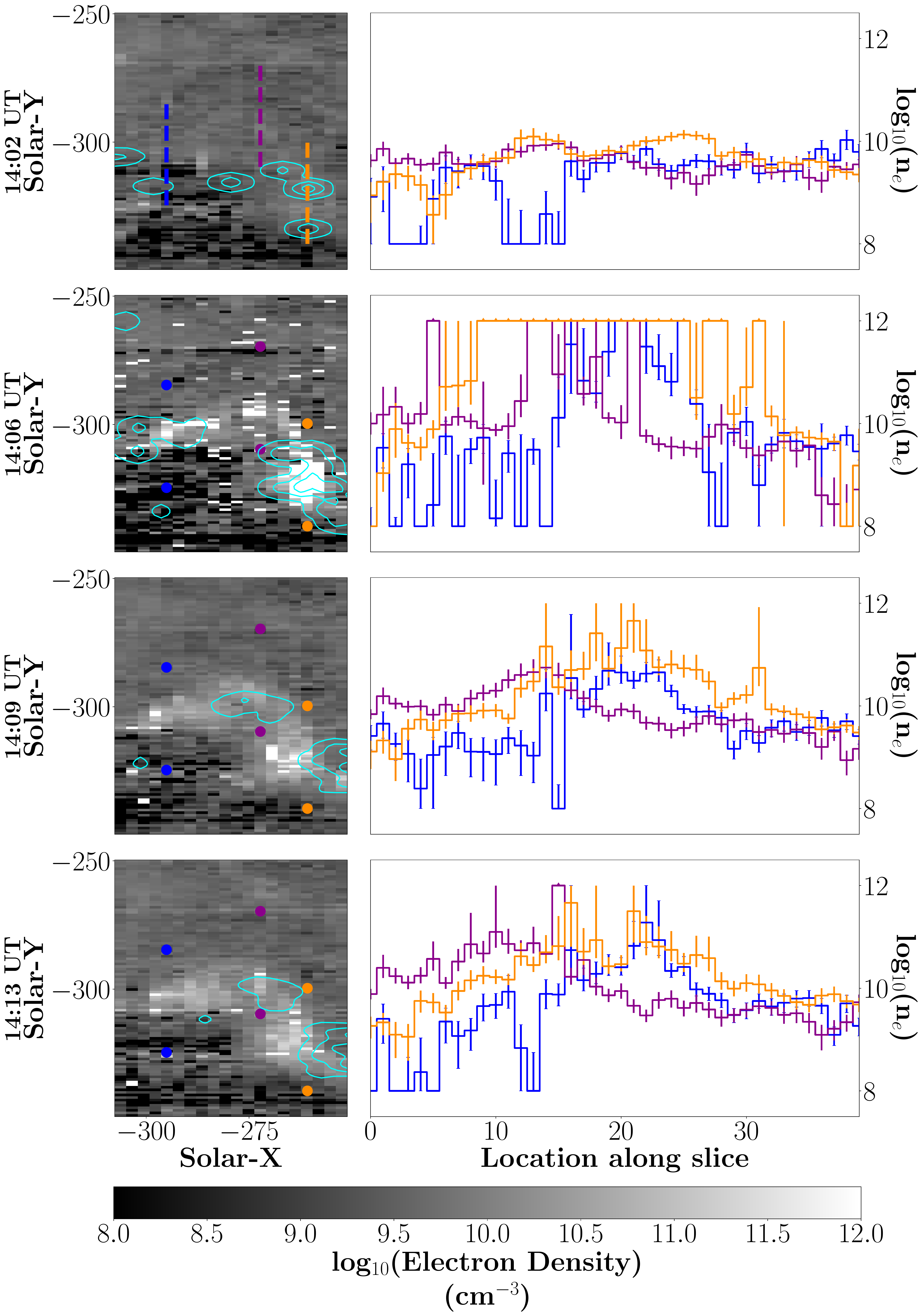}
    \caption{\ion{Fe}{14} 264.81/274.23 \AA\ calculated electron density. 
    \textbf{Left:} Density maps for four \textit{EIS} rasters spanning the impulsive, peak, and early gradual phases of the flare. 
    \textbf{Right:} Selected density slices along the flare ribbon. Each slice is cospatial with an HXR emitting source during one or more raster.
    The raster beginning at 14:02:39~UT has three dashed lines overplotted corresponding to the location in the raster of each slice. 
    The lines are omitted for following times, but the endpoints remain as a reference. 
    \textit{RHESSI} 25--50~keV CLEAN imaging contours are overlaid in cyan on all images, corresponding to 5\%, 30\%, and 70\% of the HXR maximum.}
    \label{fig:eis_dens}
\end{figure}

\begin{table}
    \centering
    \caption{Density, Velocity Correlation Coefficients}
    \begin{tabular}{p{0.15\linewidth}  p{0.18\linewidth}  p{0.235\linewidth} p{0.235\linewidth}}
    \toprule
          & \textbf{Raster:} & \textbf{Pearson $\lvert r \lvert$:} & \textbf{Pearson $\lvert r \lvert$:} \\
        \textbf{Ion:} & (UT) & n$_e$, v$_{nth}$ & v$_{nth}$, v$_{Dopp}$ \\
        \midrule
        \ion{Fe}{12} & 14:02:39 & \textbf{---} & 0.151 \\
        \ion{Fe}{12} & 14:06:13 & \textbf{---} & 0.624 \\
        \ion{Fe}{12} & 14:09:48 & \textbf{---} & 0.502 \\
        \ion{Fe}{12} & 14:13:22 & \textbf{---} & 0.526 \\
        \midrule
        \ion{Fe}{14}$^a$ & 14:02:39 & 0.037 & 0.377 \\
        \ion{Fe}{14} & 14:06:13 & 0.069 & 0.360 \\
        \ion{Fe}{14} & 14:09:48 & 0.095 & 0.114 \\
        \ion{Fe}{14} & 14:13:22 & 0.040 & 0.032 \\
        \midrule
        \ion{Fe}{14}$^b$ & 14:02:39 & 0.014 & 0.287 \\
        \ion{Fe}{14} & 14:06:13 & 0.092 & 0.063 \\
        \ion{Fe}{14} & 14:09:48 & 0.114 & 0.045 \\
        \ion{Fe}{14} & 14:13:22 & 0.112 & 0.058 \\
        \midrule
        \ion{Fe}{15} & 14:02:39 & \textbf{---} & 0.710 \\
        \ion{Fe}{15} & 14:06:13 & \textbf{---} & 0.102 \\
        \ion{Fe}{15} & 14:09:48 & \textbf{---} & 0.089 \\
        \ion{Fe}{15} & 14:13:22 & \textbf{---} & 0.273 \\
        \midrule
        \ion{Fe}{16} & 14:02:39 & \textbf{---} & 0.714 \\
        \ion{Fe}{16} & 14:06:13 & \textbf{---} & 0.506 \\
        \ion{Fe}{16} & 14:09:48 & \textbf{---} & 0.579 \\
        \ion{Fe}{16} & 14:13:22 & \textbf{---} & 0.545 \\
        \bottomrule
    \end{tabular}
    \footnotesize{ $^a$ 264.81\AA, $^b$ 274.23\AA }
    \label{tab:dens_pears}
\end{table}

Density maps formed from the \ion{Fe}{14} 264.81/274.23 \AA\ line pair are presented in Figure~\ref{fig:eis_dens} (left column) for the same four raster times as shown in Figures~\ref{fig:eis_cool_ions}~and~\ref{fig:eis_hot_ions}, with extracted slices in the Solar-Y direction shown in the right column, in order to provide a density cross section of the flare ribbon. The three slices selected for plotting are the same at every time, and are color-coordinated (such that the purple points on the left image denote the start and end of the purple curve right). The density evolution along the flare ribbon (identified by cyan \textit{RHESSI} \verb|CLEAN| contours) exceeded the upper limit of the line ratio at various times. Several regions within the ribbon exceed the limits of the intensity ratio, reaching electron densities greater than $10^{12}$~cm$^{-3}$, with the highest densities over the largest areas found in the 14:06:13~UT raster. \cite{Lee2017} focused on a particular kernel of density enhancement, the peak of which coincided with the SXR emission peak, with only a smaller enhancement found at 14:06:13~UT. However, when the entire field of view is considered, the density enhancement is greatest during the peak of the nonthermal electron event, with much of the field exceeding the limits of the density relation.
\par
Potential mechanisms responsible for excess line broadening within the flare ribbon can be investigated by correlations of density, nonthermal velocity, and Doppler velocity. A strong correlation between Doppler and nonthermal velocity within the flare ribbon may be indicative of a blend of unresolved plasma flows. Conversely, a stronger correlation between electron density and nonthermal velocity would indicate other effects, such as opacity, pressure, or potentially even turbulent broadening, are dominant. Measured correlations between these quantities within the flare ribbon are presented in Table~\ref{tab:dens_pears} for \ion{Fe}{12}, \ion{Fe}{14}, \ion{Fe}{15}, and \ion{Fe}{16}, which span a 1.15~MK range. No correlations are presented with density for \ion{Fe}{12}, \ion{Fe}{15}, or \ion{Fe}{16}, as there are no reliable density measurements in these lines.
\par
For the entire duration studied, neither \ion{Fe}{14} line exhibited any correlation between electron density and nonthermal velocity. There is a weak correlation between nonthermal velocity and the Doppler shift of the line core, with a peak correlation of $\lvert r \lvert$=0.377 in \ion{Fe}{14} 264.81\AA\ during the early flare 14:02:39~UT raster.
The two hotter lines exhibited correlation between nonthermal velocity and Doppler velocity during the 14:02:39~UT raster. By the 14:06:13~UT raster, this correlation is found only in \ion{Fe}{16}. The cooler \ion{Fe}{12} 195.12\AA\ line only exhibits correlation between nonthermal and Doppler velocities after the peak of energy injection. During the 14:06:13~UT raster, coincident with the  peak of nonthermal electron injeciton, this correlation peaked at $\lvert r \lvert = $~0.624, indicating significant unresolved flow structure.
\par
The behaviour of the \ion{Fe}{14} line pair stands in contrast with \cite{Milligan2011}, who found a strong correlation between nonthermal velocities and densities within the this line pair. The low correlations are more consistent with the findings of \cite{Doschek2007}, who studied plasma in a quiescent active region and also found no evidence of such a correlation.
\par
These correlations, taken from temperatures surrounding the FRT are a signature of explosive chromospheric evaporation, as observed in the vicinity of a major energy deposition layer. At temperatures above and below the FRT, the nonthermal widths are likely due to a superposition of unresolved flows. Near the FRT, both nonthermal and Doppler velocities were small, implying that the Doppler velocity structure was well resolved.

\subsection{IRIS Results}\label{sec:iris_results}

\begin{figure*}
    \centering
    \includegraphics[width = 18cm]{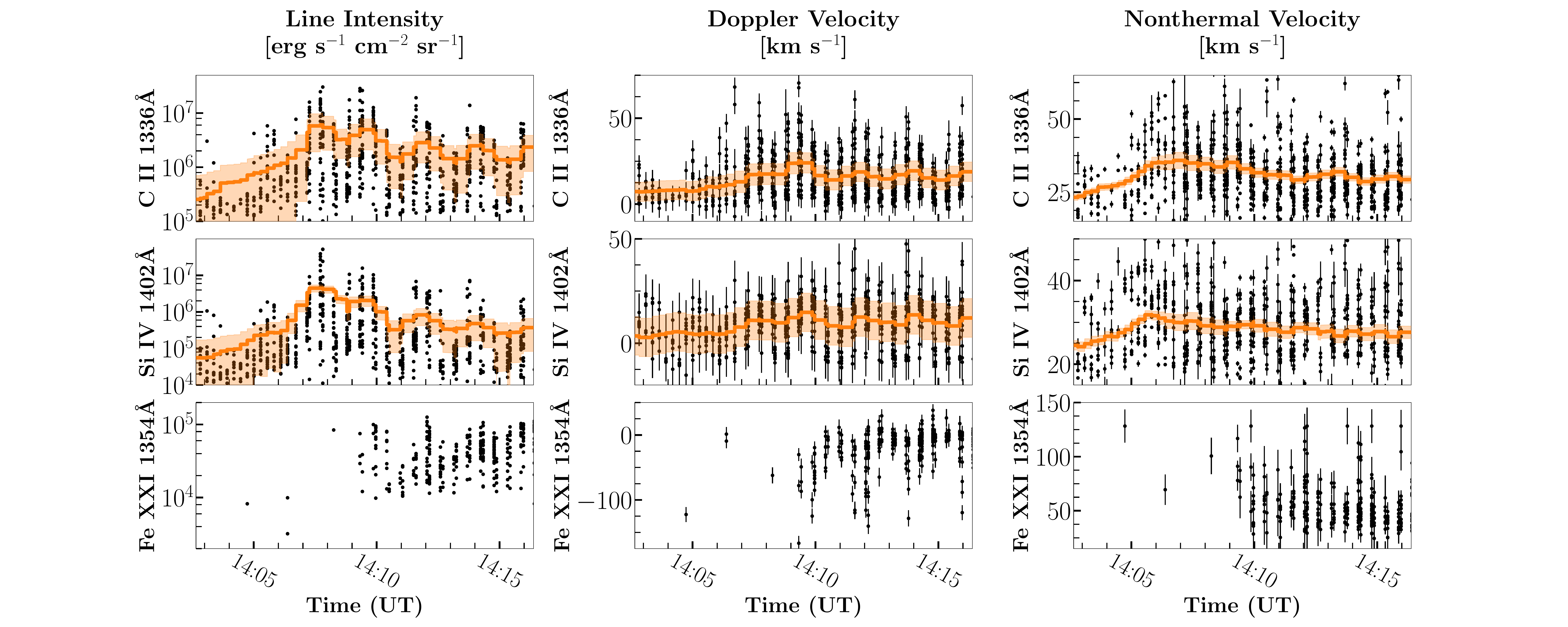}
    \caption{\textit{IRIS} spectral fitting results for the \ion{C}{2} 1335.71\AA\ line (\textbf{Top} row), \ion{Si}{4} 1402.81\AA\ (\textbf{Middle} row), and \ion{Fe}{21} 1354.07\AA\ (\textbf{Bottom} row).
    \textbf{Left Column:} Intensity for each of the three lines.
    \textbf{Middle Column:} Doppler velocity for each of the three lines.
    \textbf{Right Column:} Nonthermal velocity width for each of the three lines. For \ion{C}{2} 1335.71\AA\ and \ion{Si}{4} 1402.81\AA, which are signficantly brighter, and more easily fit that \ion{Fe}{21} 1354.07\AA , the running mean of each parameter is overlaid in orange. 
    Note that the appearance of periodicity in the running mean of \ion{C}{2} and \ion{Si}{4} parameters is an artifact induced by the loss of spatial information.}
    \label{fig:iris_vs}
\end{figure*}

\begin{figure*}
    \centering
    \includegraphics[width = 17.5cm]{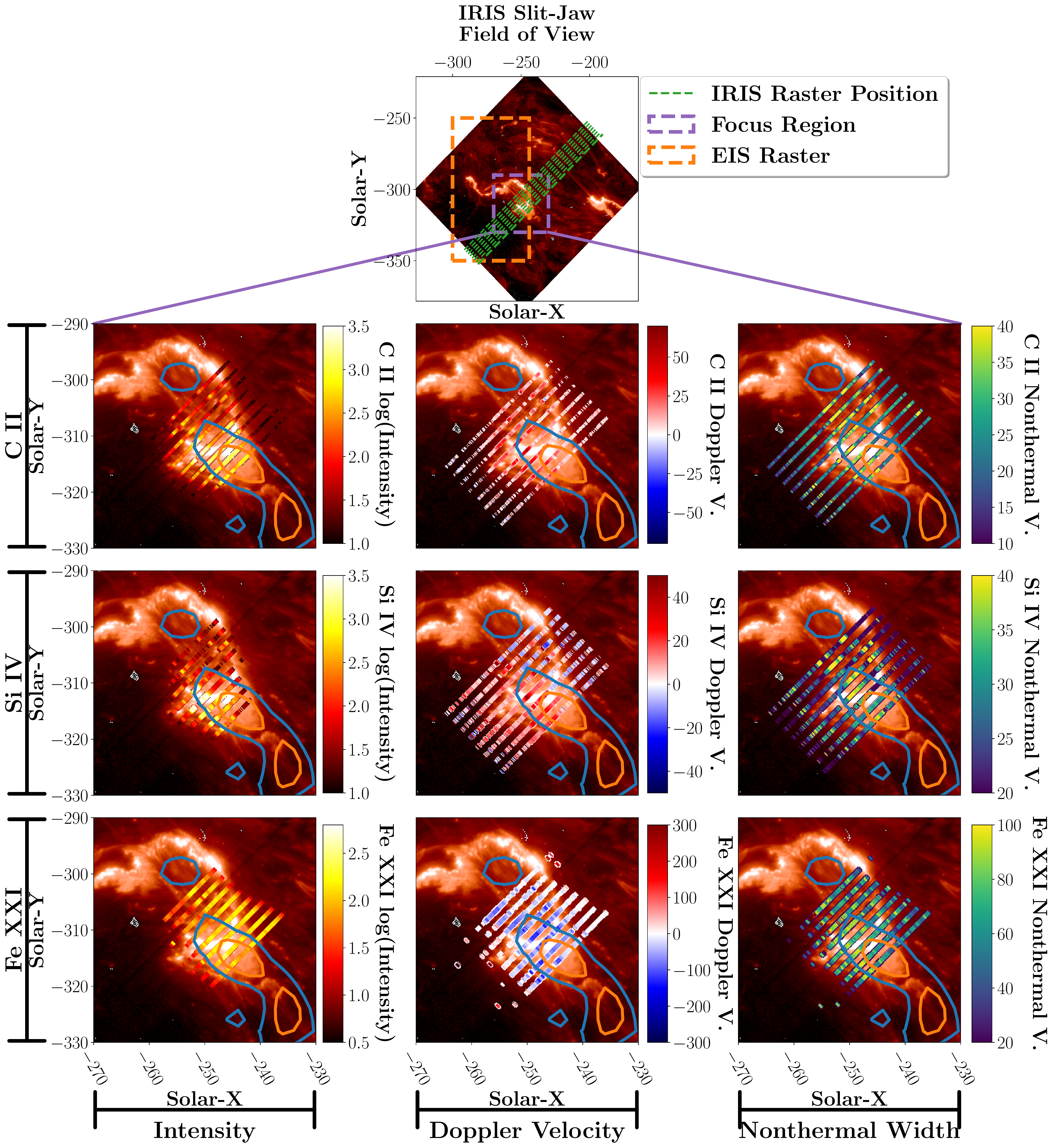}
    \caption{Spectral fitting parameters from the \textit{IRIS} raster spanning 14:08:21--14:10:24~UT, superimposed upon the co-temporal \ion{C}{2} slit-jaw image. 
    \textbf{Top:} the full slit-jaw image with the field-of-view of the \textit{EIS} instrument overlaid in orange, and each raster slit position shown in green. The purple box denotes the region of interest shown in each panel below.
    Each column denotes the line intensity (in photons~s$^{-1}$~\AA$^{-1}$), the Doppler velocity of the line (in km~s$^{-1}$, and the nonthermal width (in km~s$^{-1}$), for \ion{C}{2} 1335.71\AA\ (\textbf{Top}), \ion{Si}{4} 1402.81\AA\ (\textbf{Middle}), and \ion{Fe}{21} 1354.07\AA\ (\textbf{Bottom}). In the case of \ion{Fe}{21}, where fewer profiles could be fit, the overlaid images are integrated through the duration studied, taking the value with the larger magnitude where multiple values existed. The width of the \textit{IRIS} slit has been exaggerated by a factor of five. Overlaid contours represent the 25\% (blue) and 50\% (orange) levels of \textit{RHESSI} 40--100~keV emission.}
    \label{fig:IRIS_context}
\end{figure*}

Line intensities, Doppler velocities, and nonthermal velocities from \textit{IRIS} spectral fitting are shown in Figure~\ref{fig:iris_vs} for each of the three ions fit for pixels lying within the flare ribbon. Points lying outside the flare ribbon were masked.  The left column shows the integrated line intensity for \ion{C}{2}~1335.71\AA, \ion{Si}{4}~1402.81\AA, and \ion{Fe}{21}~1354.07\AA, while the middle shows the Doppler velocity, and the right shows nonthermal velocities. For the bright \ion{C}{2} and \ion{Si}{4} lines, the running mean of each parameter is overlaid in orange, with the 1$\sigma$ error in the running mean overlaid as filled contours in the same color.
\par
The cool \ion{Si}{4} and \ion{C}{2} ions, exhibit small Doppler shifts. Over the duration of the event, 81\% of \ion{Si}{4} profiles, and 96\% of \ion{C}{2} profiles were redshifted, with peak velocities of 47.9~$\pm$~9.6~km~s$^{-1}$ at 14:09:19~UT and 59.6~$\pm$~5.6~km~s$^{-1}$ at 14:07:16~UT, respectively. This cool choromospheric condensation provides context for \textit{EIS} observations of \ion{He}{2}. For example, \ion{He}{2} exhibited a maximum redshift of 41.7~$\pm$~5.5~km~s$^{-1}$ during the 14:06:13~UT raster. At this time (14:06:42~UT), \ion{Si}{4} redshifts peaked at 27.6~$\pm$~9.4~km~s$^{-1}$.
\par
More notable is the behaviour of the calculated nonthermal velocity for \ion{Si}{4}. The running mean of this quantity peaks at 14:05:49~UT, with a mean nonthermal velocity of 31.9$\pm$1.0~km~s$^{-1}$. This is coincident with the time of the hardest electron distribution, with a power-law index less than 6. As the nonthermal velocities in \ion{Si}{4} level off later in the flare, and finally flattens at 14:10~UT, the power-law index increases, until the nonthermal electron event ceases shortly before 14:11~UT. In the case of \ion{Si}{4}, at least, the excess widths calculated from spectral fitting may be linked to line opacity changes, driven by the deposition of energy from a particularly hard distribution of nonthermal electrons.
\par
For the hot, low-emissivity \ion{Fe}{21} line, there are comparatively few spectra with significant observable emission, particularly at earlier times. The earliest instance of an \ion{Fe}{21} profile that could be reasonably fit was at 14:04:44~UT, and was already highly blueshifted to -122.5~$\pm$~11.6~km~s$^{-1}$, with a nonthermal width of 128.2~$\pm$~15.4~km~s$^{-1}$. Most of the emission from this line during the flare impulsive phase was obscured by high levels of noise in the continuum as a consequence of shorter exposure times. At later times in the flare, \ion{Fe}{21} was observed to have Doppler velocities mostly between 0-- -80~km~s$^{-1}$, with outliers observed in excess of -150~km~s$^{-1}$ ($\lvert v_{max} \lvert = 166.67 \pm 11.4$~km~s$^{-1}$ at 14:09:19~UT). 
\par
The Doppler shifts presented here are observed earlier and have values in excess of those profiles fit by \cite{Lee2017}, who found no \ion{Fe}{21} Doppler velocities in excess of -60~km~s$^{-1}$, which they measured at 14:10~UT, for a particular kernel of emission. \cite{Li2015} were able to fit velocities as early as 13:45~UT. However, their measured Doppler velocities were, overall, smaller. Comparable Doppler velocities were found by \cite{OtherLi2015}, who studied an X1.0 flare that occurred on 2014 March 29, and found \ion{Fe}{21} Doppler velocities of $-214$~km~s$^{-1}$. \cite{Tian2015} also found similar blueshifts for the X1.6 flare on 2014 September 10, reaching a maximum of $-240$~km~s$^{-1}$, while \cite{graham2015} found velocities of up to $-300$~km~s$^{-1}$ for the same event.
\par
\ion{Fe}{21} nonthermal velocities were high for the entire duration of the flare, with a mean of 54.5~km~s$^{-1}$ and a maximum nonthermal velocity of 128.2~$\pm$~15.1~km~s$^{-1}$. These measurements are larger by than other studies of this flare. \cite{Lee2017} found no nonthermal velocities greater than $\approx 54$~km~s$^{-1}$ (0.6\AA\ FWHM) within the kernel chosen by that study. The nonthermal velocities presented here are some of the highest observed for this ion, comparable to observations by \cite{graham2015}, \cite{Polito2015}, and \cite{Polito2016}.
\par
All parameters in Figure~\ref{fig:iris_vs} exhibited a large amount of scatter. As only pixels within the flare ribbon were selected, the remaining scatter must be due to differences across the field-of-view. The spatial context for these measurements is shown in Figure~\ref{fig:IRIS_context}, which shows the line intensities, Doppler velocities, and nonthermal velocities along each raster. For the \ion{C}{2} and \ion{Si}{4} lines, the raster beginning at 14:08:21~UT was selected for the high spatial coverage and low levels of saturation. For the weak \ion{Fe}{21} line, the entire time span was stacked to provide a coherent depiction of the region of interest. Where multiple \ion{Fe}{21} profiles were present, the parameter of greatest magnitude was selected for display.
\par
When displayed in this manner, it is apparent that enhancements in \ion{C}{2} and \ion{Si}{4} intensity, Doppler velocity, and nonthermal velocity track the structure of the flare ribbon. The \ion{Fe}{21} intensities, Doppler velocities, and nonthermal velocities, however, do not appear to track the flare ribbon. Rather, enhancements in these parameters appear to trace the edges of a loop structure connecting the two flare ribbons visible in slit-jaw imaging. A similar structure appears in the hottest \textit{EIS} ions (\ion{Fe}{23} and \ion{Fe}{24}) during the 14:09:48~UT and 14:13:22~UT rasters. As this structure is not visible in any other \textit{EIS} emission lines, the minimum temperature of this structure must be between 6.31~MK (\ion{Ca}{17}) and 11.48~MK (\ion{Fe}{21}).

\subsection{Evolution of Doppler and Nonthermal Velocity as a function of Temperature}\label{sec:vel_evol_temp}

\begin{figure}
    \centering
    \includegraphics[width = \columnwidth]{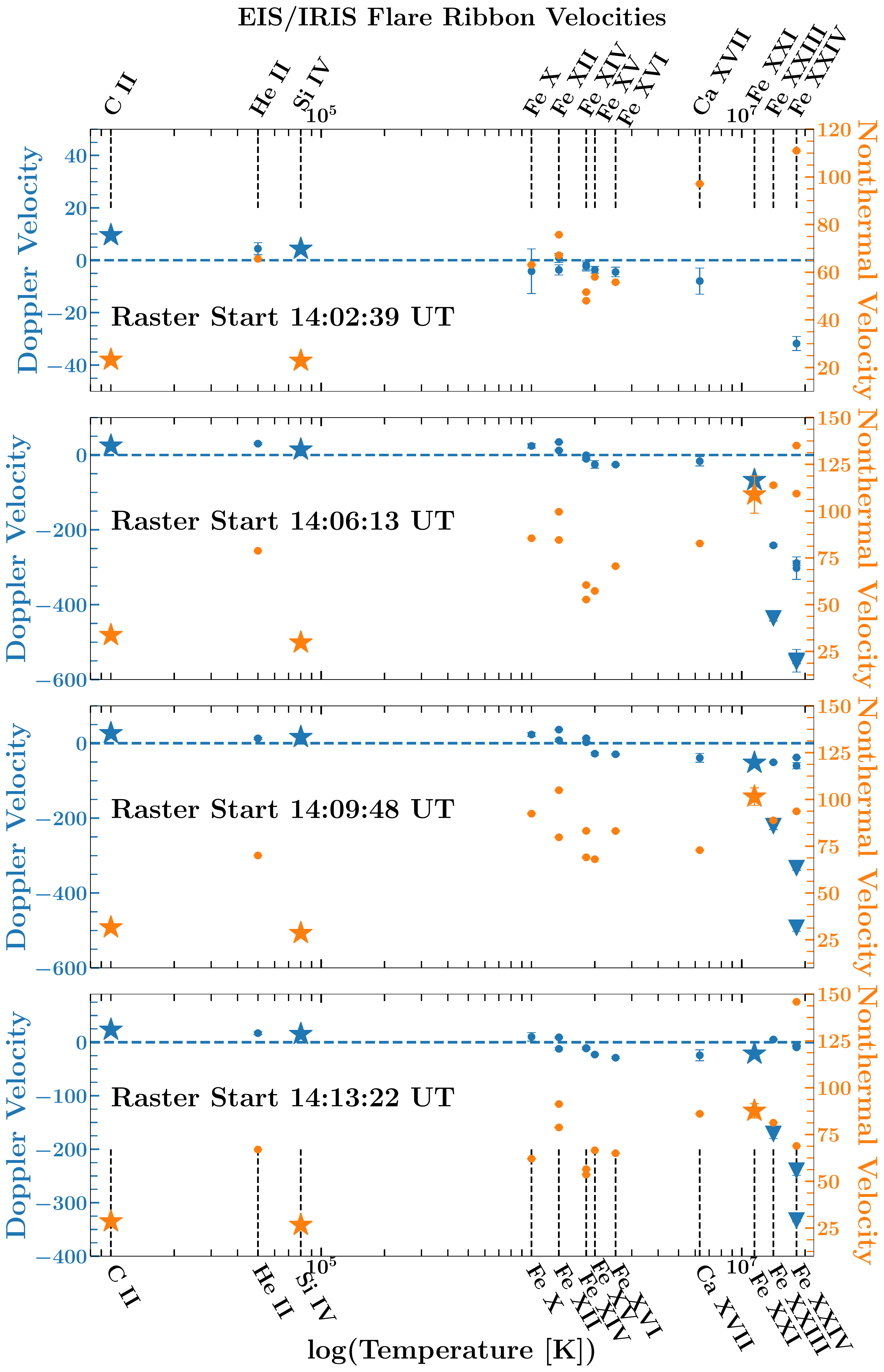}
    \caption{Spatially-averaged Doppler and nonthermal velocities taken from a region in the center of the HXR footpoint (approximate coordinates are X: -250$\arcsec$, Y: -320$\arcsec$), for each of the four rasters presented in Figure~\ref{fig:eis_cool_ions}. Doppler velocities are shown in blue, while nonthermal velocities are shown in orange. Points obtained from \textit{IRIS} spectral fitting are denoted by stars ($\star$), and points corresponding to \textit{EIS} line core fits are denoted by circles, while fit blue wing enhancements are denoted by triangles ($\nabla$). Each point displays associated error bars. However, the wide span of velocities causes many of the error bars to fall within the area subtended by the data point.}
    \label{fig:eis_ftpt}
\end{figure}
Figure~\ref{fig:eis_ftpt} shows Doppler and nonthermal velocities as a function of temperature and time for a region within the primary flare ribbon. The \textit{IRIS} data are cospatial with \textit{EIS} data, and approximately co-temporal to the extent that the differing cadences could be matched. For ions that exhibited a strong blue-wing enhancement (\ion{Fe}{23} and \ion{Fe}{24}), the Doppler velocities of the core and wing components are included, as well as the nonthermal velocity of the core component only.
\par
The FRT was clear between \ion{Fe}{12} and \ion{Fe}{14}. \textit{IRIS} observations showed redshifts from chromospheric condensation, which continued to the coolest temperatures studied. Within the blueshifted lines, the \ion{Fe}{21} line observed with \textit{IRIS} appeared to more consistent with the blueshifts observed in the cores of the \ion{Fe}{23} and \ion{Fe}{24} lines, rather than the blue wing enhancements (which were noticeable outliers in Doppler velocity).
\par
Nonthermal velocities increased with temperature from the cool \textit{IRIS} lines through the \ion{Fe}{12} line observed by \textit{EIS}. There was a sudden drop in nonthermal velocity at this temperature, observed to some extent in all time bins studied. Within the blueshifted lines, the nonthermal velocity again increased with increasing temperature approximately linearly through \ion{Fe}{24}. The \ion{Fe}{21} \textit{IRIS} line fit well into this linear relation. By the 14:13:22~UT raster, while the break in nonthermal velocities was still present, the relation has become a great deal more shallow when compared to the peak raster at 14:06:13~UT.

\section{Discussion and Conclusions}\label{sec:Conclusions}

In this work, a notably-complete set of observations were used to relate the flare-driven nonthermal energy release with the response of the chromosphere. The results presented here place an emphasis on the time-resolved profile of nonthermal electron-driven emission in conjunction with the evolving chromosphere. The nonthermal electron distribution was provided via \textit{RHESSI} spectral fitting, while emission lines observed with the \textit{EIS} and \textit{IRIS} instruments probed the response of the event in intensity, Doppler velocity, nonthermal velocity, and density. As the nonthermal electron event began and proceeded, the chromosphere was observed to transition from gentle to explosive chromospheric evaporation, with densities and high-temperature velocities peaking during the interval identified as the peak of nonthermal electron energy deposition. 
\par
Solar flares are true multiwavelength events in every sense of the word, with telltale signatures across spectral bands from radio to HXR. Events covered with a wide range of instrumentation across a wide spectral range are exceedingly rare  \citep{MilliganIreland}. A holistic understanding of the generation, transport, and deposition of flare energies may be composed by integration of the many spectral windows provided by numerous instruments in the current state-of-the-art. These connected observations of the response of the chromosphere to the call of electron injection are critical to initializing models and guiding the results of numerical simulations.
\par
The injection of nonthermal electrons lasted 352 seconds and deposited more than 4.8~$\times 10^{30}$~erg into the chromosphere. Prior to the onset of nonthermal emission, gentle chromospheric evaporation was observed in \textit{EIS} rasters, characterized by compact blueshifted regions observed in ions with T~$\geq$~1.35~MK. After this time, the chromosphere responded explosively, with upflows in excess of -50~km~s$^{-1}$ in \ion{Fe}{16}, -65~km~s$^{-1}$ in \ion{Ca}{17}, and a core blueshift of $-242$~km~s$^{-1}$ in the \ion{Fe}{24} line. 
\par
During the period of explosive chromospheric evaporation, several unique behaviours were observed in \textit{EIS} rasters. Most notable was the monolithic shift of the \ion{Fe}{24} complex. Typically, the \ion{Fe}{24} line can be well characterized by a stationary core, with a strong enhancement to the blue wing, characterized by a blend of Gaussian profiles. While this behaviour is observed at various times during the event, \textit{EIS} raster covering the peak of the flare (14:06:13 UT) exhibited monolithic shifts of the entire \ion{Fe}{24} line complex, with little to no stationary emission. This behaviour is greatly diminished by the start of the next raster, and absent by the following.
\par
The presence of blue wing enhanced spectral lines at hot temperatures was first noted in observations of \ion{Ca}{19} using the Bragg Crystal Spectrometer (BCS) aboard \textit{Yokoh} by \cite{Doschek2005}. \cite{Milligan2009} found similar profiles in \textit{EIS} observations of \ion{Fe}{23} and \ion{Fe}{24} lines during a solar flare. This behaviour was theorized to be a consequence of the low spatial resolutions of these instruments (BCS in particular was a disk-integrated instrument). The low resolution had the effect of superimposing stationary looptop emission with blueshifted footpoint emission.
Confirmation seemingly came with observations of the \ion{Fe}{21} line utilizing the higher-resolution \textit{IRIS} instrument. \cite{graham2015,Polito2015,Polito2016} found that this line exhibited no notable asymmetry. \cite{Doschek2013} and \cite{Brosius2013} both found instances of symmetric, blueshifted \ion{Fe}{23} profiles in an M1.8 and C1 flare, respectively. The behaviour exhibited by the \ion{Fe}{24} line here, where the core of the line was found to be highly blueshifted while maintaining an enhanced blue wing is not an expected behaviour.
\par
This behavior may be attributed to a superposition of unresolved flows. During the peak of this flare, several atmospheric strata with temperatures $\geq 14.1$~MK could have formed. However, the absence of a stationary population of 14.1~MK plasma until late in the flare remains unexplained. That it is present later in the event could indicate either that the stationary plasma was heated beyond the 18.2~MK \ion{Fe}{24} formation temperature, or that the looptop heating lagged behind the heating of the flare footpoint. \textit{RHESSI} spectral fitting showed the presence of plasma as hot as 70~MK during the peak of the flare, and as hot as 40~MK by the time a strong stationary core was observed at 14:13:22~UT.
\par
The temperature sampling provided by the \textit{EIS} instrument allowed constraints on the FRT, which was found to be in the range 1.35--1.82~MK. This is comparable to the FRT presented in \cite{Milligan2009}, between 1.5--2.0~MK, despite the differences in flare size (\textit{GOES} C1.1 versus X1.6). This is similar as well to values presented by several other studies, including \cite{Graham2011} (1.25--1.6~MK for a C6.6 flare), \cite{Young2013} (1.1--1.6~MK for an M1.1 flare), and \cite{Watanabe2020}, who found two FRTs; T$<$1.3~MK in one region, 1.3$<$T$<$1.8~MK in another during an X1.8 flare. \cite{Brannon2014}, however, modeled flow reversal properties in flares driven by thermal conduction, and found FRTs ranging from 0.526--4.78~MK, with some evolution in time. While the FRT range found for this event is similar to the range found in much smaller events, the area affected by the energy input is significantly larger, with a second flare ribbon well outside the \textit{EIS} field for this event. It may be that flow reversal always, or nearly always occurs around this temperature, which is independent of deposited energy.
\par
Every emission line studied exhibited line broadening. In \textit{EIS} rasters, the smallest nonthermal velocities are found just above the FRT in the \ion{Fe}{14} emission line pair. The nonthermal velocities of \textit{EIS} emission lines increase up to the FRT, with a sudden drop in nonthermal velocity just above temperature, before increasing again to the highest temperatures. Two particular emission lines, \ion{Si}{4} and \ion{Fe}{21}, both observed by the \textit{IRIS} instrument, are of note. The \ion{Fe}{21} line exhibited broad, symmetric profiles, that were often low-intensity. While the magnitude of the nonthermal widths of these profiles are not unprecedented \citep{young2015,Lee2017,Kerr2020}, they are among the broadest yet observed \citep{graham2015,Polito2015,Polito2016}. Broad, highly-shifted profiles in this line appeared early in the flare, prior to the peak of electron injection, implying that even relatively weak electron precipitation is sufficient to generate profiles with large nonthermal widths, lending further questions as to their generation \citep{Polito2019}. The cool \ion{Si}{4} line also exhibited enhanced nonthermal widths, albeit at a much lower level. These enhancements are notable due to their similarity with the evolution of the nonthermal electron spectral index, implying that the nonthermal velocity enhancement at cool temperatures may be linked directly to the deposition of energy in the lower atmosphere by nonthermal electrons.
\par
The electron density within the flare footpoint, as measured by the \ion{Fe}{14} 264.81/274.23 \AA\ ratio increased by nearly two orders of magnitude in the minutes following the onset of the electron injection event. Enhancements in nonthermal velocity in \ion{Fe}{14} were found to be small and not correlated with the density or Doppler velocity, standing in contrast to the findings of \cite{Milligan2011}. The \ion{Fe}{16} emission line exhibited correlation between Doppler and nonthermal velocity, in agreement with the findings of \cite{Milligan2011} and \cite{Doschek2013} for this emission line. A significant correlation was also observed between the Doppler and nonthermal \ion{Fe}{12} velocities, suggesting that nonthermal velocities in lines formed above and below the FRT originated from unresolved velocity flow structures along the line of sight, similar to the findings of \cite{Young2013}.
\par
This study combined temporally, spatially and spectrally-resolved, observations for a large number of distinct emission lines. This set of flare parameters combined a time-dependant electron injection profile with a time-dependant chromospheric response, including Doppler and nonthermal velocities, electron densities, and emission line intensities, with multiple rasters covering the nonthermal electron event. In addition to providing a detailed profile of this large solar flare, the derived parameters can be used to guide and interpret modeling of the atmosphere, using state-of-the-art hydrodynamic flare simulation codes, such as HYDRAD \citep{HYDRAD1,HYDRAD2}, RADYN \citep{RADYN1,RADYN2,RADYN3}, or FLARIX \citep{FLARIX1,FLARIX2}. Time-dependant parameters of nonthermal electron energy injection from \textit{RHESSI} would be used to provide the electron beam input. The chromospheric response across temperatures from 10$^{4}$--10$^{7}$~K provides guidance for the correlation of simulation outputs. Together, this allows for both a deeper understanding of the dynamic response of the chromosphere to an impulsive injection of energy, as well as the ability to constrain the numerical simulations to the underlying physics.
\par
Several unanswered questions remain: the specifics of energy and mass transport in the post-flare atmosphere as driven by relatively-weak nonthermal electron heating; the origin and nature of symmetrical nonthermal-broadened emission line profiles, especially in the extreme case of \ion{Fe}{21}; the origin of the oft-observed blue wing asymmetry observed in the hottest \ion{Fe}{23} and \ion{Fe}{24} emission lines; the unexpected shift of the typically-stationary \ion{Fe}{24} line core; and the transition from gentle to explosive chromospheric evaporation. Such questions can be answered in part by detailed simulations of coupled datasets containing both the coronal call and the chromospheric response. In the era of multi-wavelength solar observations, coordinated observations of large solar flares have become not only a possibility but a necessity. As activity increases during the rise of Solar Cycle 25, a new suite of instrumentation will enable similar studies to be conducted in unprecedented detail. Joining the venerable \textit{IRIS} and \textit{EIS} instruments, whose capabilities have not yet been exhausted, are the new \textit{STIX} \citep{STIX} and \textit{SPICE} \citep{SPICE} instruments, aboard Solar Orbiter \citep{SolarOrbiter}, which will enable a new era of observations, both in terms of their capabilities, but also their unique heliocentric positioning, which will be vital for studying the activity that will accompany the rise of this new solar cycle.

\begin{acknowledgments}
We thank the referee for the careful reading of our manuscript, as well as the extremely productive and helpful comments, which have improved the quality of the work immensely. This work was funded by NASA grant NNX17AD31G and support from DoD Research and Education Program for Historically Black Colleges and Universities and Minority-Serving Institutions (HBCU/MI) Basic Research Funding Opportunity Announcement W911NF-17-S-0010, Proposal Number: 72536-RT-REP, Agreement Number: W911NF-18-1-0484. ROM would like to thank Science and Technologies Facilities Council (UK) for the award of an Ernest Rutherford Fellowship (ST/N004981/2).
\end{acknowledgments}

\bibliography{call_and_response_0803.bib}{}
\bibliographystyle{aasjournal}
\end{document}